\documentclass[
    aps,
    prb,
    superscriptaddress,
    amsmath, amssymb,
    twocolumn,
    longbibliography,
    10pt,
    floatfix
]{revtex4-2}

\usepackage[american]{babel}
\usepackage{microtype}
\usepackage{csquotes}
\usepackage{graphicx, xcolor}
\usepackage{adjustbox}
\usepackage{dsfont}
\usepackage{mathtools}
\usepackage[useregional]{datetime2}

\usepackage[
    colorlinks,
    bookmarks=true,
    citecolor=blue,
    linkcolor=blue,
    urlcolor=blue
]{hyperref}
\usepackage{hyperxmp}
\hypersetup{
  pdftitle    = {Liouvillian topology and nonreciprocal dynamics in open Floquet chains},
  pdfauthor   = {Florian Koch, Yu-Min Hu, Jan Carl Budich},
  pdfkeywords = {Open quantum system, Floquet system, non-Hermitian topology, Lindblad master equation, nonreciprocal transport},
  pdfcreator  = {pdflatex}, 
  pdfproducer = {LaTeX with hyperref}
}

\newcommand{\U}{\mathcal{U}}
\renewcommand{\L}{\mathcal{L}}
\newcommand{\F}{\mathcal{F}}
\newcommand{\W}{\mathcal{W}}

\newcommand{\dd}{\mathop{}\!\mathrm{d}}
\DeclareMathOperator{\tr}{Tr}
\DeclareMathOperator{\hc}{h.c.}

\newcommand{\pdv}[1]{\frac{\partial}{\partial #1}}
\newcommand{\dv}[1]{\frac{\mathrm{d}}{\mathrm{d} #1}}
\newcommand{\comm}[2]{\left[#1,\,#2\right]}
\newcommand{\acomm}[2]{\left\{#1,\,#2\right\}}
\newcommand{\braket}[2]{\langle #1 | #2 \rangle}

\DeclarePairedDelimiter{\bra}{\langle}{|}
\DeclarePairedDelimiter{\ket}{|}{\rangle}
\DeclarePairedDelimiter{\dket}{|}{\rangle\!\rangle}
\DeclarePairedDelimiter{\dbra}{\langle\!\langle}{|}
\DeclarePairedDelimiter{\expval}{\langle}{\rangle}

\begin{document}


\title{Liouvillian topology and nonreciprocal dynamics in open Floquet chains}


\newcommand{\TUD}{Institute of Theoretical Physics, Technische Universität Dresden and W\"urzburg-Dresden Cluster of Excellence ctd.qmat, 01062 Dresden, Germany}
\newcommand{\MPIPKS}{Max Planck Institute for the Physics of Complex Systems, N\"othnitzer Str.~38, 01187 Dresden, Germany}

\author{Florian Koch}
\affiliation{\MPIPKS}
\affiliation{\TUD}

\author{Yu-Min Hu}
\affiliation{\MPIPKS}

\author{Jan Carl Budich}
\email{jcbudich@pks.mpg.de}
\affiliation{\MPIPKS}
\affiliation{\TUD}

\date{\DTMdisplaydate{2026}{06}{22}{-1}}

\begin{abstract}
    Open quantum systems far from thermal equilibrium can exhibit remarkable physical phenomena including topological properties without a direct equilibrium counterpart.
    Along these lines, in periodically driven-dissipative systems within the effective non-Hermitian (NH) Hamiltonian approximation spectral winding numbers have been linked to intriguing nonreciprocal transport properties.
    Here, going beyond an NH Hamiltonian description, we introduce and study a microscopic lattice model of a driven open quantum system described by a Markovian quantum master equation, which exhibits the mentioned spectral winding within an NH approximation.
    By encompassing quantum jump processes in the topological analysis, we uncover a distinct \emph{jump-induced} topological phase, which qualitatively corresponds to the richer nonreciprocal transport properties of the fully quantum model.
    In addition, we find that the NH skin effect, i.e., the accumulation of a macroscopic number of eigenstates at one end of the system, is already visible in the transient dynamics even for systems with periodic boundary conditions.
    Our results exemplify the subtle correspondence between NH topological properties and physical manifestations of Liouvillian topological properties in open quantum systems, thus providing a theoretical framework towards understanding unidirectional transport in quantum dissipative Floquet dynamics.\newline
    \newline
    DOI: \href{https://doi.org/10.1103/284g-926d}{10.1103/284g-926d}
\end{abstract}

\maketitle

\section{Introduction}
Exploring the role of topology in non-equilibrium dynamics has led to the discovery of numerous intriguing topological phenomena, some of which do not have a direct equilibrium counterpart \cite{Kitagawa2010, Jiang2011, Kitagawa2012, Rudner2013, Vajna2015, Budich2016, Martin2017, Quelle2017, Flaschner2018, Junk2020}.
Two interdisciplinary frontiers of current research along these lines are provided by the study of periodically driven (Floquet) systems \cite{Lindner2011, Gomez-Leon2013, Cayssol2013, Potter2016, Eckardt2017, Yao2017, Oka2019} and the effective non-Hermitian (NH) Hamiltonian approach to dissipative dynamics \cite{Gong2018, Zhou2019, Kawabata2019, Budich2020, Bergholtz2021, Lin2023, Okuma2023}.
Bridging the two aforementioned concepts, topological properties of effective NH Floquet Hamiltonians describing driven dissipative systems have recently been investigated \cite{Gong2015, Longhi2017, Hockendorf2020, Zhang2020a, Wu2021, Sierant2022, Zhou2023, Koch2024, Dash2024, Wanjura2025}.
This raises the natural question as to what extent these findings may be transferred from the effective NH Hamiltonian level to a fully quantum mechanical description in the framework of quantum master equations \cite{Song2019, Minganti2020, Mori2023, Monkman2026}.
Motivated by preliminary insights into the relation of NH and Liouvillian topology \cite{Dai2016, Budich2017, Koch2022, Koch2024, Chaduteau2025, Chen2025}, the objective of our present work is to address this question both at the abstract level of defining the quantum dissipative counterpart of NH Floquet topological invariants and at the practical level of predicting observable physical phenomena relating to them.

\begin{figure}[htbp]
    \includegraphics[width=0.99\columnwidth]{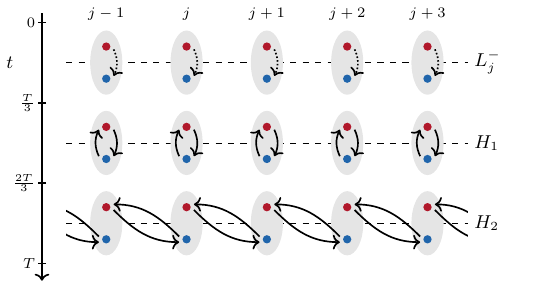}
    \caption{
        Schematic illustration of the dissipative Floquet chain of spin-$\tfrac{1}{2}$ fermions over one Floquet period $T$.
        Each site $j$ (gray circle) hosts spin-up (red) and spin-down (blue) states.
        Dotted arrows denote spontaneous decay via $L_j^-(t) = \sqrt{\gamma(t)} \sigma_j^-$, while solid arrows indicate coherent dynamics generated by the Hamiltonians $H_1$ and $H_2$.
        This serves as the minimal model studied here, alongside more experimentally realistic variants discussed in the text.
    }
    \label{fig:model}
\end{figure}

Below, we introduce a microscopic quantum model for a dissipative one-dimensional (1D) Floquet chain by adding on-site spin relaxation to a helical Floquet channel (see Fig.~\ref{fig:model} for an illustration), which is characterized by the locking of spin and direction of motion \cite{Kane2005,Budich2017}.
When employing an effective NH Hamiltonian approximation, the classical counterpart of our model realizes an NH topological Floquet phase with complex spectral winding of the Floquet operator in the sense of Ref.~\cite{Hockendorf2020}.
Interestingly, we find that considering the full Liouvillian dynamics governed by the Lindblad master equation including quantum jump processes significantly enriches the phenomenology.
In particular, the preferred direction of nonreciprocal transport \cite{Liu2022,Geng2023}, which arises from the spin-momentum locking of the clean model due to dissipation, becomes tunable with basic model parameters.
This finding is in qualitative agreement with a sign change of a Liouvillian winding number that we construct as a topological invariant for the quantum model and which is absent in the approximate NH description.
Moreover, we demonstrate how the NH skin effect, i.e., the accumulation of a macroscopic number of eigenstates at one end of a 1D chain with open boundary conditions  \cite{Lee2016, Yao2018, Kunst2018, Yokomizo2019, Lee2019, Okuma2023}, leaves clear fingerprints in the (subextensive) transient dynamics even in the case of periodic boundary conditions.
Our results are corroborated by the numerically exact solution of the quantum master equation.

The remainder of this article is structured as follows.
In Sec.~\ref{sec:Model}, we introduce the quantum model as a dissipative extension of the helical Floquet model of Ref.~\cite{Budich2017}, where the dissipation is engineered to act only during an additional Floquet waiting phase, serving as a toy model to illustrate the key concepts in contrast to the physical dissipation treated later.
In Sec.~\ref{sec:NH_approx}, we investigate the effective non-Hermitian approximation, and in Sec.~\ref{sec:open_quantum_system} we turn to the full dissipative quantum dynamics governed by the Lindblad master equation.
Each of these sections is structured into three parts: (1) the topological properties; (2) transport and topology under periodic boundary conditions; and (3) the behavior under open boundary conditions, including the non-Hermitian skin effect.
Finally, in Sec.~\ref{sec:experimental_realization}, we propose an experimentally feasible implementation that incorporates dissipation (spontaneous decay and dephasing) throughout the dynamics.

\section{Model}
\label{sec:Model}
We consider a periodically driven one-dimensional lattice model of spin-$1/2$ fermions with dissipation, whose dynamics is governed by the time-dependent Lindblad master equation \cite{Lindblad1976, Gorini1976}
\begin{align}
    \dv{t}\rho = \L(t)[\rho] = -i\comm{H_\mathrm{S}(t)}{\rho} + \sum_\kappa \gamma_\kappa(t) \mathcal{D}_{L_\kappa}[\rho] \label{eq:Lindblad}
\end{align}
where $\rho$ is the density matrix, $H_\mathrm{S}(t)$ is the periodically driven Hamiltonian, $\gamma_\kappa(t)$ are time-periodic dissipation rates, and $\mathcal{D}_{L_\kappa}[\rho] = L _\kappa \rho L_\kappa^\dagger- \frac{1}{2} \acomm{L_\kappa^\dagger L_\kappa}{\rho}$ is the Lindblad dissipator for the jump operators $L_\kappa$.
As introduced in the following, the Floquet driving is implemented through a linearly segmented three-phase cycle of total duration $T$, such that
\begin{align}
    H_\mathrm{S}(t+T) = H_\mathrm{S}(t), &  & \gamma_\kappa(t+T) = \gamma_\kappa(t).
\end{align}

The spinor operator $\psi_j = (\psi_{j\downarrow}, \psi_{j\uparrow})$ annihilates a fermion on the lattice with unit lattice constant on the $j$th site.
With the standard Pauli matrices $\sigma_{x,y,z}$ \cite{Pauli1927} and the spin-flip operators $\sigma_{\pm} = \frac{1}{2}(\sigma_x \pm i\sigma_y)$, we define the Hamiltonians $H_1 = \beta\sum_{j=1}^L \psi_j^\dagger \sigma_x \psi_j$ and  $H_2 = -\alpha \sum_{j=1}^L \psi_j^\dagger \sigma_+ \psi_{j+1} + \hc$.

We also consider the spontaneous decay of the spin that acts locally on each site $j$, which is described by the jump operators \cite{Breuer2009,Gardiner2010}
\begin{align}
    L_j^- = \psi_j^\dagger \sigma_- \psi_j. \label{eqn:Ljs}
\end{align}
We further assume that the dissipation rates are equal on each site, i.e., $\forall j: \gamma_j(t) = \gamma(t)$.
The time dependence of the Hamiltonian and the dissipation rate can then be written as
\begin{align}
    H_\mathrm{S}(t) & = \begin{cases}
                            0   & \text{if } t \in [0, \frac{T}{3}),             \\
                            H_1 & \text{if } t \in [\frac{T}{3},  \frac{2T}{3}), \\
                            H_2 & \text{if } t \in [\frac{2T}{3}, T).
                        \end{cases} \\
    \gamma(t)       & = \begin{cases}
                            \gamma & \text{if } t \in [0, \frac{T}{3}), \\
                            0      & \text{if } t \in [\frac{T}{3}, T).
                        \end{cases}
\end{align}
In Sec.~\ref{sec:experimental_realization}, we extend this model to a more realistic setting by including dephasing and applying dissipation during the full time evolution.
Throughout the article, we set the parameter $\alpha = \frac{3\pi}{2T}$.
Further, we denote site indices by $m, m', n, n', j$ and spin indices by $s, s', r, r'$.
For brevity and readability, we use the shorthand $\sum_{s\in\{\downarrow, \uparrow\}} \to \sum_s$ and $\sum_{j=1}^L \to \sum_j$ (and similarly for the other indices introduced above).

Defining the particle number operator $N = \sum_{j, s} \psi_{js}^\dagger \psi_{js}$ and the translation operator $T_a = \sum_{j,s}\psi_{j+a,s}^\dagger\psi_{j,s}$, we find that at all times $t$, the dynamics conserve particle number,
\begin{align}
    \comm{H_\mathrm{S}(t)}{N} & = \comm{L_j^-}{N} = 0, \label{eq:number_conservation_Liouvillian}
\end{align}
and is translation invariant (assuming periodic boundaries $\psi_{L+a,\sigma} = \psi_{a,\sigma}$)
\begin{align}
    T_1 \L(t)[\rho] T_1^\dagger & = \L(t)[T_1 \rho T_1^\dagger] \label{eq:translation_invariance_Liouvillian}.
\end{align}
We stress that our present discussion focuses on the single-particle sector of the model.
The corresponding single-particle states are defined by $\ket{j,s} = \psi_{j,s}^\dagger \ket{\text{vac}}$.
It follows that the Liouvillian propagator $\U(t_0, t) = \mathcal{T} \exp(\int_{t_0}^t \L(t') \dd t')$ is also number conserving and translation invariant.
We define the Floquet-Liouvillian propagator
\begin{align}
    \F = \U(T, 0),
\end{align}
which describes the time evolution over a single Floquet period and thus governs the stroboscopic dynamics.

For (Hermitian) Hamiltonians, one typically uses the Floquet Hamiltonian $H_\mathrm{F}$ defined by $\exp(-i H_\mathrm{F} T) = U(T, 0)$, where $U(T, 0)$ is the time evolution operator over a single period, to analyze the stroboscopic dynamics.
However, since they are related by a logarithm, the spectral properties of the Floquet Hamiltonian and the one-period propagator are closely connected -- in particular, topological quantities defined for the Floquet Hamiltonian can be equivalently defined for the one-period propagator.
In the Liouvillian case, however, the effective Floquet-Liouvillian $\L_\mathrm{F}$ defined by $\exp(\L_\mathrm{F} T) = \F$ is not as useful, since it is not necessarily a generator in the form of a time-independent Lindblad master equation and thus does not have a clear physical interpretation \cite{Schnell2020}.
Therefore, we will focus on the Floquet-Liouvillian propagator $\F$ in the following discussion, which is also compatible with the discussion of Floquet non-Hermitian topology in Ref.~\cite{Hockendorf2020}.

In the limit of zero dissipation, this model recovers the Helical Floquet model of Ref.~\cite{Budich2017}.
Since the Hamiltonians are translational invariant, we can apply a Fourier transform $\ket{k,s} = \frac{1}{\sqrt{L}}\sum_j e^{-ijk} \ket{j,s}$ and write the unitary time evolution operator $U^k(T, 0) = \exp(-iH_2^k T/3) \exp(-i H_1^k T / 3)$ in one period $T$ in lattice momentum space, where $H_1^k = \beta\sigma_x$ and $H_2^k = -\alpha[\cos(k)\sigma_x - \sin(k)\sigma_y]$ are the Bloch Hamiltonians corresponding to $H_1$ and $H_2$, respectively.
A particularly interesting special case occurs for $\frac{\alpha T}{3} = \frac{\beta T}{3} = \frac{\pi}{2}$, where $U^k(T, 0) = e^{ik\sigma_z}$ realizes perfect helical spin transport \cite{Budich2017}; i.e., as $k$ varies from $-\pi$ to $+\pi$, the two eigenvalue branches wind in opposite direction -- one clockwise and the other counterclockwise -- around the unit circle.
In the following, we address an important question: How does the dissipation alter the transport in the original closed system.
As a first step, we will take a look at the effective non-Hermitian approximation where a theory of topological transport has been developed in Ref.~\cite{Hockendorf2020}.

\section{Effective non-Hermitian approximation}
\label{sec:NH_approx}
The effective non-Hermitian Hamiltonian is defined as \cite{Ashida2020,Griffiths1993,Wiseman1996,Daley2014}
\begin{align}
    H_\mathrm{NH}(t) = H_\mathrm{S}(t)  - \frac{i}{2} \sum_\kappa \gamma_\kappa(t)  L_\kappa^\dagger L_\kappa. \label{eq:NH_Hamiltonian}
\end{align}
By neglecting the quantum jumps ($\propto L_\kappa \rho L_\kappa^\dagger$), the Lindblad master equation reduces to a von-Neumann-like equation \cite{Sergi2019}
\begin{align}
    \dv{t}\rho(t)  = -i (H_\mathrm{NH}(t)  \rho (t) - \rho(t)  H_\mathrm{NH}(t) ^\dagger)
\end{align}
which is solved by $\rho(t) = U_\mathrm{NH}(t, t_0) \rho(t_0) U^\dagger_\mathrm{NH}(t, t_0)$, with $U_\mathrm{NH}(t, t_0) = \mathcal{T} \exp(-i\int_{t_0}^t H_\mathrm{NH}(\tau) \dd\tau)$.
This shows that under this time evolution, a pure state remains pure, i.e.,  $\ket{\varphi(t)} = U_\mathrm{NH}(t, t_0) \ket{\varphi(t_0)}$.
Similar to the Lindbladian case, we define the one-period Floquet propagator $F = U_{\mathrm{NH}}(T,0)$, which controls the stroboscopic dynamics.
However, since $U_\mathrm{NH}(t, t_0)$ is nonunitary, the norm $\lVert\varphi(t)\rVert = \sqrt{\braket{\varphi(t)}{\varphi(t)}}$ is not conserved, reflecting gain and loss.
In open quantum systems, this change can be interpreted as the probability of a quantum jump occurring \cite{Daley2014}.
Consequently, a substantial deviation of $\braket{\varphi(t)}{\varphi(t)}$ from unity signals that quantum jumps become relevant.

We note that a complementary interpretation of an effective NH Hamiltonian approximation is to neglect quantum noise in a quantum Langevin equation \cite{Breuer2009,Gardiner2010,Koch2022}.
In this context, which is familiar in photonic systems, the NH Hamiltonian is thus tantamount to a classical limit.

In this section, we focus on the nonunitary dynamics of pure states, which is in principle applicable to either of the aforementioned interpretations of NH Hamiltonians, aiming at discussing the topology of non-Hermitian Floquet chains.
A quantum treatment based on the full Lindblad master equation will be elaborated in the next section.

\subsection{Topology of non-Hermitian Floquet chains}
\label{ssec:topology}
In Ref.~\cite{Hockendorf2020}, a topological winding number for non-Hermitian Floquet chains has been introduced and linked to transport properties of the system.
Here, a short review of this theory is given, adapting it to the NH (or classical) limit of our model with spin in addition to spatial degrees of freedom.

\begin{figure*}[htbp]
    \includegraphics[width=\textwidth]{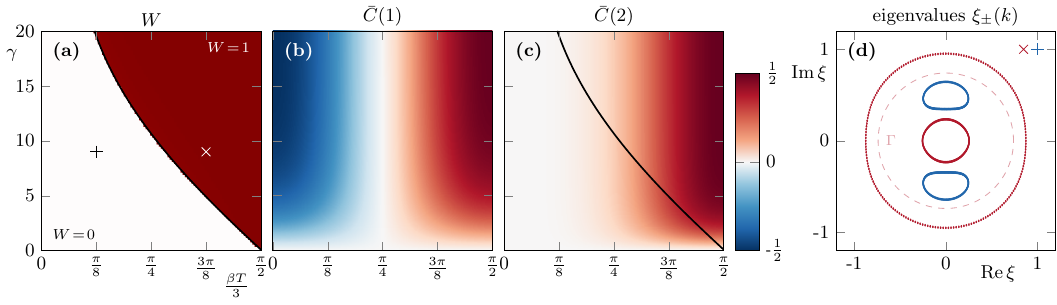}
    \caption{
        The correspondence between the topological winding number and the charge transport properties of the Floquet non-Hermitian system [Eq.~\eqref{eq:NH_Hamiltonian}] without quantum jumps.
        (a) Topological winding number as a function of the parameters $\beta$ and $\gamma$.
        The parameter sets $(\frac{\beta T}{3}, \gamma) = (\frac{\pi}{8}, 9)$ and $(\frac{3\pi}{8}, 9)$ are highlighted, with their corresponding eigenvalues shown in panel (d).
        (b), (c) Mean transferred charge $\bar C(p)$ after $p=1$ [panel (b)] and $p=2$ [panel (c)] Floquet cycles, plotted vs $\beta$ and $\gamma$.
        (d) Spectral distribution of the nonunitary Floquet propagator $F$ in the punctured complex plane; the dashed red circle shows a choice of an imaginary gap $\Gamma$, with respect to which the red spectrum becomes topologically nontrivial.
        Throughout the article, we set the parameter $\alpha = \frac{3\pi}{2T}$.
    }
    \label{fig:NH_transport}
\end{figure*}

Let the stroboscopic dynamics of the translation invariant system be described by the nonunitary Floquet operator
\begin{align}
    F =  \sum_{s,s'} \int_{-\pi}^{\pi} F^{ss'}(k) \ket{k,s}\bra{k,s'} \dd k, \label{eq:Floquet_operator_momentum}
\end{align}
the spectrum of which is restricted to the punctured complex plane $\mathds{C}\backslash \{0\}$ due to the invertibility $F(t)^{-1} = F(-t)$ of the propagator.
The individual eigenvalues $\xi_a(k) = e^{-i\epsilon_a(k)}$ of $F(k)$ therefore must form closed loops in $\mathds{C}\backslash \{0\}$ due to the periodicity of the quasienergies $\epsilon_a(k)$.
Considering the origin of the complex plane as the natural point gap for the Floquet operator, eigenvalues winding around it create a nontrivial topology.
To this end, a winding number $W(\Gamma)$ for the dominant quasienergy bands is defined as
\begin{align}
    W(\Gamma) = \frac{i}{2\pi} \sum_{e^\Gamma < |\xi_a|} \int_{-\pi}^\pi \xi_a(k)^{-1} \partial_k \xi_a(k) \dd k \label{eq:W_Höckendorf}
\end{align}
where $\Gamma$ defines an imaginary gap $i\Gamma$ that separates the quasienergies.
The condition $e^\Gamma<|\xi_a(k)|$ picks out the dominant quasienergy bands of the system.
Figure \ref{fig:NH_transport}(d) shows the eigenvalues for two parameter sets, with blue indicating topologically trivial and red indicating nontrivial phases, where an imaginary gap is given by the dashed red line.

This winding number can be related to the mean transferred charge $\bar C = \bar C_n(1)$, which is defined as the average propagation distance
\begin{align}
    \bar C_n(p) = \frac{1}{2p} \sum_s \langle\varphi_{ns}(pT) | x - n | \varphi_{ns}(pT)\rangle \label{eq:Cbar}
\end{align}
where $\ket{\varphi_{ns}(pT)} = F^p \ket{n,s}$ is the state after $p$ Floquet cycles, $x = \sum_{j,s} j \ket{j,s}\bra{j,s}$ is the position operator, and $n$ is the initial site index, which is irrelevant due to translation invariance under periodic boundaries.
The prefactor $\frac{1}{2p}$ reflects averaging over $p$ Floquet cycles and the two spin degrees of freedom.
Introducing the spin-averaged initial density matrix $\rho_n(0) = \frac{1}{2}\sum_{s} \ket{n,s}\bra{n,s}$, which describes an unpolarized fermion on site $n$, we can write the average propagation distance compactly as $\bar C_n(p) = \frac{1}{p} \expval{x - n}_{\rho_n(pT)}$, where $\rho_n(pT) = F^p \rho_n(0) (F^\dagger)^p$.
Note that the time evolved states must not be normalized because of the non-Hermiticity, and thus, the propagation distance is weighted by $\tr[\rho_n(pT)] \neq 1$.

In the thermodynamic limit $L\to\infty$ and with periodic boundaries, the mean transferred charge $\bar C$ can be expressed in the lattice momentum space as
\begin{align}
    \bar C = \frac{i}{4\pi} \int_0^{2\pi} \tr_S\left[F^\dagger(k) \partial_k F(k)\right] \dd k
\end{align}
where $\tr_S A = \sum_s A_{ss}$ denotes the trace over the spin degrees of freedom of the Fourier-transformed operator.
For regularized dynamics (RD), this gives the fundamental relation
\begin{align}
    \bar C \mathrel{\overset{\text{\tiny(RD)}}{=}} \frac{1}{2} W(\Gamma)
\end{align}
where regularized dynamics requires that (1) the dominant eigenvalues of the Floquet propagator have modulus one, (2) all other eigenvalues have modulus infinitesimally close to zero, and (3) the corresponding eigenvectors are mutually orthogonal.
The factor $\frac{1}{2}$ arises from our definition of the lattice constant: In our convention, one lattice unit corresponds to the distance between neighboring unit cells, whereas in Ref.~\cite{Hockendorf2020} it is defined as the distance between adjacent sublattice sites.
Figure \ref{fig:NH_transport}(a) shows the topological winding number as a function of the system parameters $\beta$ and $\gamma$.

\subsection{Winding number and transport}
For our model, specifically in the dissipative subperiod, the non-Hermitian Hamiltonian reads
\begin{align}
    H_\mathrm{NH} = (-i)\frac{\gamma}{4} \sum_j \psi_j^\dagger (\mathds{1} + \sigma_z) \psi_j.
\end{align}
For the special case of perfectly helical unitary dynamics ($\alpha = \beta = \frac{3\pi}{2T}$) \cite{Budich2017} in the coherent subperiods, the total nonunitary Floquet operator $F(k)$ reads
\begin{align}
    F(k) =
    \begin{pmatrix}
        e^{ik - \frac{\gamma T}{6}} & 0       \\
        0                           & e^{-ik}
    \end{pmatrix} =
    \begin{pmatrix}
        \xi_-(k) & 0        \\
        0        & \xi_+(k)
    \end{pmatrix}.
\end{align}
It leads to an imaginary splitting of the eigenvalues $\xi_\pm(k)$ due to the damping of the left-moving mode.
Furthermore, when $\gamma\to\infty$, the requirements of regularized dynamics are fulfilled, and the transport is quantized with $\bar C = \frac{1}{2} W(\frac{1}{2}) = \frac{1}{2}$.
However, this particular limit is only achieved by fully damping one of the two modes, which -- in the context of open quantum systems -- may be seen as unphysical since quantum jumps were ignored.

In Fig.~\ref{fig:NH_transport} we show the topological winding number and the mean transferred charge $\bar C(p)$ for $p=1$ and $p=2$ of the non-Hermitian Floquet chain.
To calculate the winding number, we numerically check whether the two eigenvalue bands are separated by an imaginary gap $\Gamma$ and calculate the winding number $W(\Gamma)$ [see Eq.~\eqref{eq:W_Höckendorf}].
With the onset of dissipation $\gamma > 0$, a nonzero winding number emerges at $\frac{\beta T}{3} = \frac{\pi}{2}$ as the two eigenvalue circles of the perfect-momentum locking Hamiltonian split.
As $\gamma$ increases, the nontrivial topology extends to arbitrarily small values of $\beta$.
However, for $\beta = 0$, we will never reach a topologically nontrivial phase because the $k$ dependence of the eigenvalues vanishes:
\begin{align}
    F(k) =
    \begin{pmatrix}
        0 & ie^{ik} \\ i e^{-ik - \frac{\gamma T}{6}} & 0
    \end{pmatrix}\ \Rightarrow \
    \xi_\pm(k) = \pm i e^{-\frac{\gamma T}{12}}
\end{align}
In contrast to the system’s topological properties, the mean transferred charge after a single Floquet cycle [cf.~\ref{fig:NH_transport}(b)] can point in either direction depending on the parameters, while still being nonreciprocal in each case.
This behavior can be traced back to the microdynamics: The spin-down component of the initial state remains unaffected by dissipation, whereas the spin-up component is partially depleted, and then rotated by $H_1$ and transported via $H_2$.
The resulting asymmetry induces nonreciprocal motion.
However, the leftward transport is merely a transient one-cycle effect.
After this first cycle, the part of the state transferred to the left is in a spin-up configuration and is subsequently removed by the system’s non-Hermiticity [cf.~Fig.~\ref{fig:NH_transport}(c)].
For $\frac{\beta T}{3} \neq \frac{\pi}{2}$, the rightward transport also gradually diminishes, since the spin-down component is slightly rotated by the non-optimal $\beta$, producing a spin-up fraction that is subsequently removed by dissipation.
Consequently, under the effective non-Hermitian time evolution, the initial spin-down component -- and thus the sustained nonreciprocal transport -- can only persist for the optimal value of $\beta$.

In summary, for our model within the topological phase, nonreciprocal transport proceeds from the second Floquet period onward in the direction dictated by the topological winding number.
Quantization of the transported charge occurs only in the limit of the regularized dynamics.
Nevertheless, the presence of a topological phase does not guarantee stable transport; this is ensured only in the regularized limit.
From the perspective of the quantum master equation, however, this limit should be viewed critically, since it relies on the idealized assumption that quantum jumps can be neglected.

\subsection{Non-Hermitian skin effect}
With open boundaries, nonreciprocal transport may lead to an accumulation of the quantum state on one of the boundaries.
The accumulation of eigenstates of the Hamiltonian at the edge is called the non-Hermitian skin effect (NHSE)  \cite{Yao2018}.
In order to investigate the extent to which our system exhibits NHSE and how this relates to the topological winding number, we consider the system with open boundary conditions in this section.
In particular, we use the normalized (right) eigenstates $\ket{\varphi_a}$ of the nonunitary Floquet operator $F$ to investigate their localization $\bra{\varphi_a} x \ket{\varphi_a}$.

\begin{figure}
    \includegraphics[width=0.99\columnwidth]{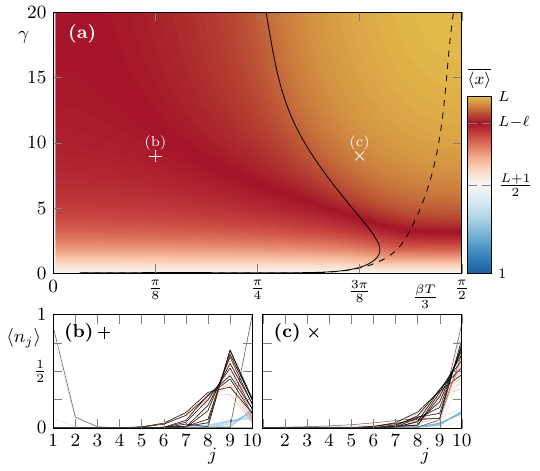}
    \caption{
    Eigenstate localization of the nonunitary Floquet operator $F$ under open boundary conditions for a chain of length $L=10$, which shows the existence of non-Hermitian skin effect.
    (a) Mean localization $\overline{\langle x \rangle} = \frac{1}{2L}\sum_{a=1}^{2L}\langle \varphi_a| x |\varphi_a\rangle$ of the eigenstates $\ket{\varphi_a}$ as a function of the system parameters $\beta$ and $\gamma$.
    The dashed line indicates the phase boundary, where two edge states with vanishing imaginary part of the eigenvalues emerge, and the solid line indicates the phase boundary, where one of the two edge states $\ket{\varphi_\mathrm{L}}$ localizes at the left, i.e., $\langle\varphi_\mathrm{L}|x|\varphi_\mathrm{L}\rangle < \frac{L+1}{2}$.
    Both phenomena occur to the left of their respective phase boundaries.
    The parameter sets $(\tfrac{\beta T}{3}, \gamma) = (\tfrac{\pi}{8}, 9)$ and $(\tfrac{3\pi}{8}, 9)$ are highlighted, with their corresponding right eigenstates shown in panels (b) and (c).
    The colorbar is chosen to explicitly highlight the limiting values of the mean localization.
    In the limits $\gamma\to\infty$ and $\beta\to 0_+$, we find $\overline{\langle x \rangle} \to L-\ell = L - \frac{3}{2}(1 - \frac{1}{L})$, while for $\beta \to \frac{3\pi}{2T}$ we obtain $\overline{\langle x \rangle} \to L$.
    (b), (c) Spatial distributions $\langle \varphi_a| n_j |\varphi_a\rangle$ (black) together with spin-resolved components $\langle \varphi_a|n_{js}|\varphi_a\rangle$ for $s=\uparrow$ (red) and $s=\downarrow$ (blue).
    There exists a single eigenstate localized at the left edge of the chain in panel (b).
    }
    \label{fig:NH_skin_effect}
\end{figure}

In Fig.~\ref{fig:NH_skin_effect}, we show the mean localization  $\overline{\langle x \rangle} = \frac{1}{2L}\sum_{a=1}^{2L}\langle \varphi_a| x |\varphi_a\rangle$ [Fig.~\ref{fig:NH_skin_effect}(a)], as well as the states of two specific parameter choices for $\beta$ and $\gamma$ [Figs.~\ref{fig:NH_skin_effect}(b) and \ref{fig:NH_skin_effect}(c)].
From the mean localization, one can see that as $\gamma$ increases, an extensive number of states accumulates on the right side of the chain.
In fact, for optimal $\frac{\beta T}{3} \to \frac{\pi}{2}$ and strong dissipation $\gamma$, every eigenstate of $F$ localizes on the right boundary.
Interestingly, when $\beta \to 0_+$, there emerge two isolated edge modes with vanishing imaginary part (the spectrum of $F$ under open boundary conditions is not shown here), one of which is localized on the right boundary and the other moves to the left boundary, as demonstrated by Fig.~\ref{fig:NH_skin_effect}(b).
The remaining modes localize on the penultimate site on the right-hand side of the chain, which is a consequence of the interplay between the absence of the spin-backrotation mechanism (as $\beta \to 0$) and the non-Hermitian part of the spontaneous decay channel.
Thus, in the limit of $\gamma \to \infty$ and $\beta \to 0_+$, the mean localization $\overline{\langle x \rangle}$ approaches $L - \frac{3}{2}(1 - \frac{1}{L})$ in contrast to $\gamma \to \infty$ and $\beta \to \frac{3\pi}{2T}$, where $\overline{\langle x \rangle} \to L$.
The emergence of these two edge modes is indicated by the dashed line in Fig.~\ref{fig:NH_skin_effect}(a) and the phase boundary, where one ($\ket{\varphi_\mathrm{L}}$) of the two modes localizes at the left boundary ($\langle \varphi_\mathrm{L}|x|\varphi_\mathrm{L}\rangle < \frac{L+1}{2}$) is indicated by the solid line.

The above numerical results imply that the nonzero winding number of the spectrum with periodic boundary conditions is closely related to the localization of the isolated edge modes with vanishing imaginary part, but not necessarily to the existence of the extensive number of states localized at the right boundary.
The localization of bulk modes resembles the non-Hermitian skin effect in static systems, which originates from a nontrivial point-gap topology that is defined slightly differently from Eq.~\eqref{eq:W_Höckendorf} \cite{Okuma2020,Zhang2020}.

Note that the inverse participation ratio also shows the localization of the eigenstates, but it does not distinguish between localization at the left or right boundary and thus fails to show the localization of the single eigenstate at the left boundary in Fig.~\ref{fig:NH_skin_effect}(b).

\section{Open quantum system}
\label{sec:open_quantum_system}
In the following, we will investigate the open quantum system including quantum jump terms in Eq.~\eqref{eq:Lindblad}.
To this end, we introduce a Fourier transform that block diagonalizes the vectorized Lindblad-Floquet operator allowing to generalize the idea of a winding number to translation invariant Liouvillians.
Afterward, we investigate the dynamics of the open quantum system for closed boundaries and the steady-state for open boundaries.

\begin{figure*}[htbp]
    \includegraphics[width=\textwidth]{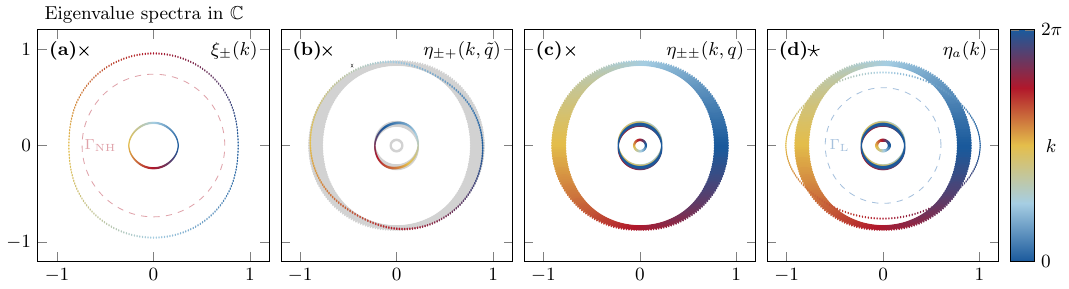}
    \caption{
    Spectral distributions in the punctured complex plane and their dependence on momentum $k$.
    (a) $\xi_\pm(k)$ of the nonunitary Floquet operator $F(k)$; the dashed red circle shows a choice of an imaginary gap $\Gamma_\mathrm{NH}$, with respect to which the spectrum becomes topologically nontrivial.
    (b) $\eta_{\pm +}(k, \tilde q) = \xi_\pm(\tilde q-k)\xi_+^*(\tilde q)$ from the no-jump Floquet-Liouvillian propagator $\F^{\mathrm{NH}}(k, q) = F(q-k) \otimes F^*(q)$, shown vs $k$ for fixed $\tilde q=2\pi\frac{33}{L}$.
    The gray area shows the eigenvalues of $\F^{\mathrm{NH}}(k, q)$ for $q\in[0,2\pi)$.
    (c) $\eta_{\pm\pm}(k, q)$ as functions of $k$ for $q\in[0,2\pi)$.
    (d) $\eta_a(k)$ of the full Floquet–Liouvillian propagator $\F(k)$ for the open Floquet chain.
    The dashed blue circle shows a choice of an imaginary gap $\Gamma_\mathrm{L}$, with respect to which the spectrum becomes topologically nontrivial.
    For all plots, the parameters are $\alpha = \frac{3\pi}{2T}, \beta = \frac{3}{4}\frac{3\pi}{2T}, \gamma=9$, and $L=200$ unit cells.
    The figure shows the embedding of the non-Hermitian topology into the Liouvillian description of the system and demonstrates the effects of the inclusion of quantum jumps on the spectrum and topology of the open system.
    In this specific case, the nontrivial topology of $F$ is not altered by the quantum jumps.
    }
    \label{fig:scatter_plots}
\end{figure*}

\subsection{Embedding of non-Hermitian topology to Liouvillians}
Even though the jump operators $L_\kappa$ in the Lindblad master equation (\ref{eq:Lindblad}) act on the density matrix $\rho$ from both sides, it is a linear differential equation in the density matrix $\rho$.
Thus, we may transform the Lindblad master equation into the usual form $\dv{t} \dket{\rho(t)} = \L \dket{\rho(t)}$ of a linear equation by vectorizing the density matrix $\rho \to \dket\rho$ \cite{Horn1994}.
The Lindblad superoperator $\L$ is then represented as a square matrix $\L$ acting on the vectorized density matrix $\dket{\rho}$ from the left only.

This differential equation can be solved by matrix exponentiation if the Lindblad superoperator does not explicitly depend on time $t$.
Thus, for each Liouvillian $\L_i$ we have
\begin{align}
    \dket{\rho(t+\Delta t)} = \exp(\L_i \Delta t) \dket{\rho(t)}.
\end{align}
In particular, the Floquet-Liouvillian propagator $\F$, which describes the stroboscopic dynamics after a full Floquet cycle, reads
\begin{align}
    \F = \exp(\L_3 T / 3)  \exp(\L_2 T / 3)  \exp(\L_1 T / 3). \label{eq:Floquet_propagator}
\end{align}
As a short-hand notation for the stroboscopic time evolution of the system, we will henceforth use
\begin{align}
    \dket{\rho(t+T)} = \F \dket{\rho(t)} \Leftrightarrow \rho(t+T) = \F[\rho(t)].
\end{align}
Further, we use the notation $\rho(t+pT) = \F^p[\rho(t)]$ for $p$ Floquet steps.

A usual vectorization scheme just orders each column of the density matrix one below the other and thus does not take into account the crucial translation invariance of the system [cf.~Eq.~\ref{eq:translation_invariance_Liouvillian}].
Due to the translation invariance of the Liouvillians (and thus of the Floquet operator), there exists a momentum variable $k$ that block diagonalizes the Floquet operator.
To this end, we apply a Fourier transform along the diagonal of the density matrix
\begin{align}
    \rho_\Delta^{ss'}(k) = \frac{1}{\sqrt{L}}\sum_j e^{-ikj} \ket{j,s}\bra{j+\Delta, s'}.
\end{align}
The vectorized state is then written as $\rho_\Delta^{ss'}(k) \to \dket{k;\Delta,s,s'}$.

\begin{figure*}[htbp]
    \includegraphics[width=\textwidth]{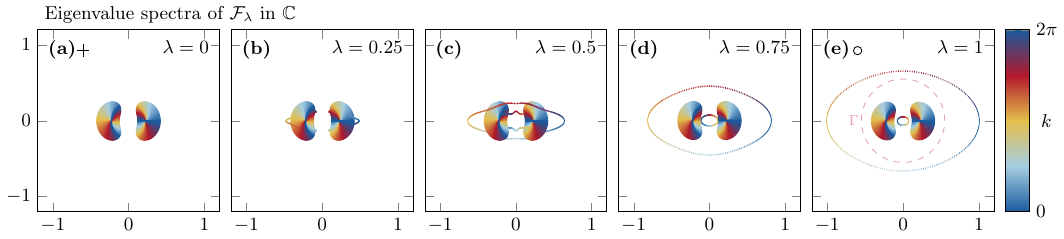}
    \caption{
        Spectral distributions of the interpolation between the no-jump Floquet-Liouvillian propagator $\F^\mathrm{NH}(k)$ and the fully open Floquet-Liouvillian propagator $\F(k)$.
        The interpolation parameter $\lambda$ is defined in the generalized quantum master equation [Eq.~\eqref{eq:lambda_Liouvillian}].
        For all plots, the parameters are $\alpha = \frac{3\pi}{2T}, \beta = \frac{1}{4}\frac{3\pi}{2T}, \gamma=9$ and $L=200$ unit cells.
        The dashed red circle in panel (e) shows a choice of an imaginary gap $\Gamma$, with respect to which the outlier spectrum becomes topologically nontrivial.
        The figure shows the effect of quantum jumps as they are gradually included in the dynamics, starting from the non-Hermitian approximation and ending with the full Lindbladian description.
        In this specific case, the quantum jumps induce a topological phase transition from a trivial to a nontrivial phase.
    }
    \label{fig:specLambda}
\end{figure*}

In this particular lattice-momentum space, the Liouvillians are diagonal with respect to $k$ and can be written as
\begin{align}
    \L_1(k)
     & = (-i\beta) \left[\mathds{1} \otimes \left(\sigma_x \oplus (-\sigma_x)\right)\right]                   \\
    \L_2(k)
     & = (i\alpha) \bigl[T_{-1} \otimes ((e^{ik} \sigma_-) \oplus (-\sigma_+)) \nonumber                      \\
     & \qquad + T_1 \otimes ((e^{-ik} \sigma_+) \oplus (- \sigma_-)) \bigr]                                   \\
    \L_-(k)
     & = \gamma \left(\L_-^\mathrm{NH} + \L_-^\mathrm{jumps}\right)                                           \\
    \L_-^\mathrm{jumps}
     & = \delta_{\Delta=0} \otimes  \sigma_- \otimes \sigma_-                                                 \\
    \L_-^\mathrm{NH}
     & = -\frac{1}{2} \left[\mathds{1}_L \otimes \left(\sigma_+\sigma_- \oplus \sigma_+\sigma_-\right)\right]
\end{align}
where $\otimes$ denotes the Kronecker product, $A \oplus B = A \otimes \mathds{1} + \mathds{1} \otimes B$ is the Kronecker sum, and $T_a$ is the translation operator.
Obviously, the Lindblad-Floquet superoperator is diagonal with respect to $k$ as well:
\begin{align}
    \F = \sum_{\nu_1,\nu_2} \F_{\nu_1\nu_2}(k) \dket{k;\nu_1}\dbra{k;\nu_2}
\end{align}
with the multi-indices $\nu_i = (\Delta_i, s_i, s_i')$.
The matrix $\F(k)$ is a $4L\times 4L$ matrix whose eigenvalues we denote by $\eta_a(k)$ with $a \in \{1, 2, \ldots, 4L\}$.
For a more detailed derivation of this equation, see Appendix \ref{app:vectorization}.

In order to get insights into the topological structure of the full Lindblad-Floquet operator, we first note that the non-Hermitian approximation [ignoring $\gamma \delta_{\Delta=0} \otimes  \sigma_- \otimes \sigma_-$ in $\L_-(k)$] is translation invariant with respect to $\Delta$ as well.
Applying another Fourier transform (momentum variable $q$) and using the properties of the Kronecker sum, we get
\begin{align}
    \F^\mathrm{NH}(k, q) = F(q-k) \otimes F^*(q) \label{eqn:F_product}
\end{align}
where $F$ is the non-Hermitian Floquet operator in momentum space [see Eq.~\eqref{eq:Floquet_operator_momentum}].
This specific structure allows to read the eigenvalues of $\F^\mathrm{NH}(k, q)$ as $\eta_{\pm\pm}(k, q) = \xi_\pm(q-k) \xi^*_\pm(q)$.
If we consider $\eta_{\pm\pm}(k, \tilde q)$ (with $q = \tilde q$ fixed) as $\xi_\pm(\tilde q-k)$ being linearly scaled with the complex number $\xi^*_\pm(\tilde q) \neq 0$, we can calculate the non-Hermitian winding number $W(\Gamma)$ by calculating the winding number of $\eta_{\pm\pm}(k, \tilde q)$.
Note that the existence of an imaginary gap $\Gamma$ separating $\xi_\pm$ forces the existence of two imaginary gaps $\Gamma_1$ and $\Gamma_2$ separating the eigenvalue sets $\{\eta_{++}\}, \{\eta_{+-},\eta_{-+}\}$ and $\{\eta_{--}\}$.
Also, the specific choice of $\tilde q$ is irrelevant for the topological properties.

Figure \ref{fig:scatter_plots} illustrates the transformation for a topologically nontrivial parameter set -- starting from the NH Floquet operator, through the Liouvillian without quantum jumps, to the full Lindbladian.
One observes that one of the two eigenvalue branches $\xi_\pm(k)$ of the NH Floquet operator $F$ [specifically $\xi_+(k)$ in Fig.~\ref{fig:scatter_plots}(a)] gives rise to two distinct eigenvalue branches $\eta_{+\pm}(k, \tilde q)$ for each value of $\tilde q$, which remain separated from each other [see Fig.~\ref{fig:scatter_plots}(b)].
The complete set of eigenvalues $\eta_{\pm\pm}(k, q)$ thus forms three distinct regions: the innermost region corresponding to $\eta_{--}(k, q)$, the middle region containing $\eta_{+-}(k, q)$ and $\eta_{-+}(k, q)$, and the outermost region corresponding to $\eta_{++}(k, q)$ [see Fig.~\ref{fig:scatter_plots}(c)].
For $\eta_{++}(k, q)$, each individual curve $\eta_{++}(k, \tilde q)$ possesses a topological winding number that is opposite in sign to that of $\xi_+(k)$.
In total, $\eta_{++}(k, q)$ consists of $L$ eigenvalue curves, each sharing the same winding number.
Accordingly, we associate the broadened eigenvalue band $\eta_{++}$ with the topological winding number of $\xi_+(k)$.
Analogously, the same correspondence holds for $\eta_{--}(k, \tilde q)$.
In contrast, for the middle region, the windings of the eigenvalue curves $\eta_{+-}(k, \tilde q)$ and $\eta_{-+}(k, \tilde q)$ are opposite and thus cancel each other, leading to a total winding number of $W = 0$.

\subsection{Topology of open quantum Floquet chains}
Including the quantum jumps, this simple relation in Eq.~\eqref{eqn:F_product} gets destroyed because the translation invariance with respect to $\Delta$ is lost.
Indeed, for our simple model, we find that for each $k$, the quantum jumps pose a low-rank perturbation to the Lindblad-Floquet propagator since it only acts on $\Delta=0$ (see Appendix \ref{app:vectorization}).
The spectrum of low rank perturbations, that is, of matrices of the form $A + \delta A$, where $\delta A$ has low rank, is an active field of mathematical research \cite{Kato1995,Benaych-Georges2011,Mehl2011,Tao2013,Bordenave2024}.
In several studied cases, the majority of eigenvalues are only weakly affected, whereas a few eigenvalues are shifted far away from the bulk spectrum.
For low-rank perturbations on a non-normal matrix, such a simple picture generally does not hold, as the entire spectrum may be shifted.
Nevertheless, even though $\F(k)$ is generally not a normal matrix, our numerical studies show that at most two outliers $o_\pm(k)$ appear for each $k$ (cf.~Figs.~\ref{fig:scatter_plots} and \ref{fig:specLambda}).

The appearance of a dominant outlier, as it can be seen in Fig.~\ref{fig:specLambda}, is demonstrated by continuously switching on jump terms within a generalized Lindblad master equation:
\begin{align}\label{eq:lambda_Liouvillian}
    \L_\lambda[\rho] = -i(H_\mathrm{NH} \rho - \rho H_\mathrm{NH}^\dagger) + \lambda \sum_\kappa \gamma_\kappa L_\kappa \rho L_\kappa^\dagger
\end{align}
where $\L_\lambda, H_\mathrm{NH}$, and $\gamma_\kappa$ formally depend on the time $t$, and the definition of the Floquet propagator $\F_\lambda$ is similar to Eq.~\eqref{eq:Floquet_propagator}.
Figure \ref{fig:specLambda} shows that, as $\lambda$ goes from 0 to 1, two outlier curves $o_\pm(k)$ appear, while the bulk spectrum remains essentially unchanged.
One of those outliers $o_+(k)$ eventually dominates the spectrum and thus the dynamics.
However, there are also parameter combinations where the bulk spectrum and outliers are comparable in magnitude as seen in Fig.~\ref{fig:scatter_plots}.

\begin{figure}[!ht]
    \includegraphics[width=\columnwidth]{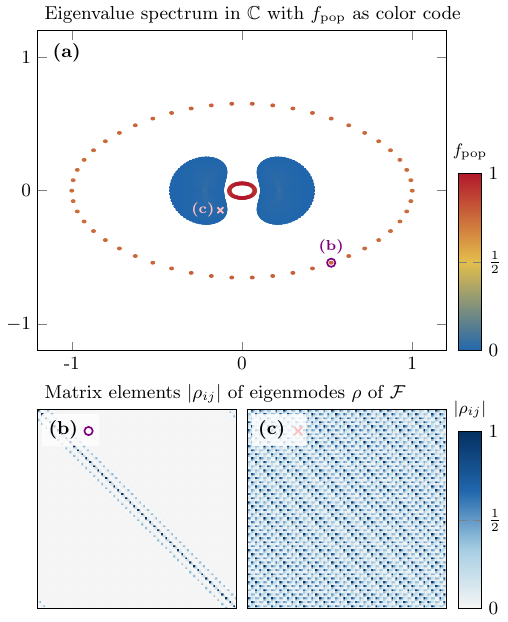}
    \caption{
        (a) Fraction of population $f_\mathrm{pop}[\rho]$ [see Eq.~\eqref{eq:f_pop}] for the eigenmodes $\rho$ of the Floquet-Liouvillian propagator $\F$.
        The parameters are $\alpha = \frac{3\pi}{2T}, \beta = \frac{1}{4}\frac{3\pi}{2T}, \gamma=9$ and $L=50$ unit cells.
        The figure shows that the outlier modes are dominated by populations, whereas the bulk modes are dominated by coherence.
        This supports the interpretation that the outlier modes are the ones that carry the transport, whereas the bulk modes represent coherence between states and thus do not directly contribute to transport.
        (b), (c) Density matrix elements (normalized to unity) of an outlier mode [panel (b)] and a bulk mode [panel (c)], which further supports the above interpretation.
        The outlier mode is dominated by diagonal elements, whereas the bulk mode is dominated by off-diagonal elements.
    }
    \label{fig:f_pop}
\end{figure}

To get a physical interpretation of the outlier modes, with $m,n$ labeling both the site and spin indices, we calculated the fraction of the population
\begin{equation}
    f_\mathrm{pop}[\rho] = \frac{\sum_m |\rho_{mm}|^2}{\sum_{m, n} |\rho_{mn}|^2} \label{eq:f_pop}
\end{equation}
of the eigenmode $\rho$ in the Fock space, which is the weight of diagonal elements in the operator norm of $\rho$.
We can similarly define the fraction of coherence
\begin{equation}
    f_\mathrm{coh}[\rho] = \frac{\sum_{m\neq n} |\rho_{mn}|^2}{\sum_{m, n} |\rho_{mn}|^2} = 1 - f_\mathrm{pop}[\rho]
\end{equation}
as the weight of off-diagonal elements.
We find (cf.~Fig.~\ref{fig:f_pop}) that these two quantities separate the bulk modes from the outlier modes, with the outlier modes dominated by diagonal parts and the bulk modes dominated by off-diagonal parts.
It means that these outlier modes can be viewed as \enquote{defect modes} caused by the low-rank perturbation $\sum_\kappa\gamma_\kappa(t)L_\kappa\rho(t) L_\kappa^\dagger$ acting only on the diagonal parts of the density matrix.
This separation becomes clearer with increasing $\gamma$ and is already very pronounced for $\gamma=9$.

Since the charge transport is detected by the expectation value of the position operator $x = \sum_{j,s} j \ket{j,s}\bra{j,s}$, a diagonal operator in the Fock space, the above eigenoperator structure supports that the outlier modes are the ones that carry the charge transport, whereas the bulk modes represent coherence between states and thus do not directly contribute to transport.

To show the relation between charge transport and the spectral topology, we define the winding number of an individual band $\eta_a(k)$ by
\begin{align}
    \W(\eta_a) = \frac{(-i)}{2\pi} \int_{-\pi}^\pi \eta_a(k)^{-1} \partial_k \eta_a(k) \dd k. \label{eq:W_individual}
\end{align}
Further, we define the system winding number $\W_\mathrm{S}$ by two distinct cases:
(1) If the outlier is separated from the bulk states by an imaginary gap, the topological winding number is the winding number of the outlier [cf.~Fig.~\ref{fig:specLambda}(e)]; and
(2) if the outlier and the bulk states are comparable in magnitude and all states have the same winding number, the topological winding number is the shared winding number of all states [cf.~Fig.~\ref{fig:scatter_plots}(d)].
Mathematically, we define the set $S_\Gamma(k) = \{\eta_a(k) : |\eta_a(k)| > \Gamma\}$ and calculating the winding number for each state within $S_\Gamma(k)$ yields another set $\W_\Gamma = \{\W(\eta_a) : \eta_a \in S_\Gamma\}$.
If $\W_\Gamma$ contains only a single winding number $\W_\Gamma = \{\W_o\}$, this must be the winding number of the outlier and we take it as the winding number of the whole system $\W_\mathrm{S} = \W_o$.
If $\W_\Gamma$ contains exactly $L$ winding numbers ($|\W_\Gamma| = L$, where $L$ is the number of unit cells) and all winding numbers are equal ($\exists \W_o \in \W_\Gamma,\ \forall \W(\eta) \in \W_\Gamma: \W_o = \W(\eta)$), then $\W_\mathrm{S} = \W_o$.
Otherwise, we set $\W_\mathrm{S} = 0$.
Note that this definition is equal to defining the winding number as the winding number of the outlier as long as the bulk states are not comparable in magnitude and winding in the opposite direction.

\subsection{Winding number and transport}
Similar to the non-Hermitian case [cf.~Eq.~\eqref{eq:Cbar}], we define the transferred charge over $p$ Floquet periods by $\bar C(p)=\bar C_n(p) = \frac{1}{p} \expval{x - n}_{\rho_n(pT)}$, which is independent of $n$ due to translational invariance under periodic boundary condition, and use the shorthand notation $\bar C = \bar C(1)$ for the single Floquet period transferred charge.
In the Fourier transformed space,  this can be written as (see Appendix~\ref{app:transferred_charge} for the derivation)
\begin{align}
    \bar C = \frac{(-i)}{2}\sum_{s,s'} \pdv{k}\F_{(0ss),\,(0s's')}(k)\Big|_{k=0}
\end{align}
One can see that the specific structure of $\bar C$ that allows the existence of the fundamental relation ($\bar C = \frac{W}{2}$) arises from the Kronecker product structure of the Floquet operator in the non-Hermitian case [$\F^\mathrm{NH} = F \otimes F^*$, see Eq.~\eqref{eqn:F_product}].
The inclusion of quantum jumps breaks this specific structure and thus the fundamental relation does not hold anymore.
However, as we will see below, the topological winding number still approximately captures the transport properties of the system.

\begin{figure}[!ht]
    \includegraphics[width=0.99\columnwidth]{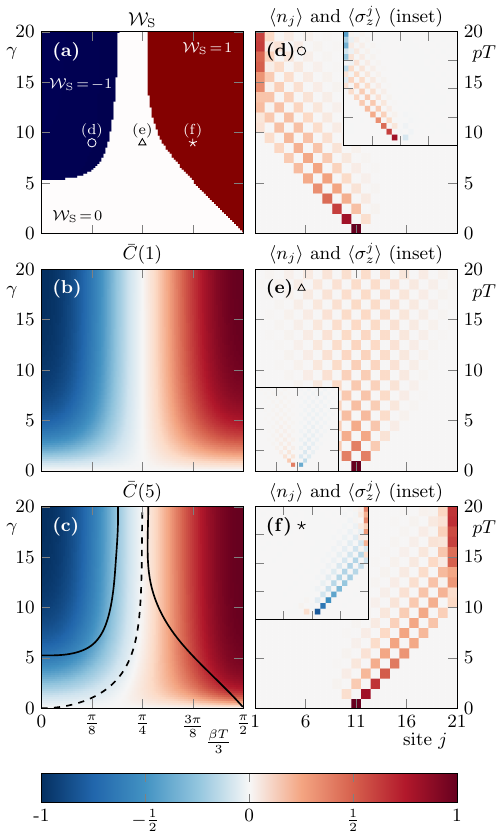}
    \caption{
        (a) Topological winding number $\W_\mathrm{S}$ as a function of the parameters $\beta$ and $\gamma$.
        (b), (c) Mean transferred charge $\bar C(p)$ after $p=1$ [panel (b)] and $p=5$ [panel (c)] Floquet cycles, plotted vs $\beta$ and $\gamma$.
        The solid lines in panel (c) indicate the contour lines of the topological phases shown in panel (a).
        The dashed line corresponds to $\gamma=-\frac{3}{T}\ln\cos(\frac{2\beta T}{3})$, which has been found to phenomenologically capture the long-time limit.
        (d)-(f) Stroboscopic time evolution of the site-resolved particle number $\expval{n_j(pT)}$, with the inset displaying the corresponding spin polarization $\expval{\sigma_z^j(pT)}$.
        The time evolution in all subplots starts from an unpolarized fermion initially placed at the center of the open chain.
        With $\gamma=9$ fixed, we set $\frac{\beta T}{3} = \frac{\pi}{8}$ [panel (d)], $\frac{\beta T}{3} = \frac{\pi}{4}$ [panel (e)], and $\frac{\beta T}{3} = \frac{3\pi}{8}$ [panel (f)].
        Panels (a)-(c) can be seen as the full Lindbladian analog of the non-Hermitian topological phase diagram shown in Figs.~\ref{fig:NH_transport}(a)-\ref{fig:NH_transport}(c), where an additional topological phase emerges around the regime $\beta=0$ and $\gamma \approx 5$.
        This additional phase can also be seen qualitatively in the transport properties.
    }
    \label{fig:local_transport}
\end{figure}

Figure \ref{fig:local_transport} shows the winding number, the transferred charge $\bar{C}(p)$ for $p=1$ and $p=5$, and the time evolution of a single unpolarized fermion placed in the middle of the dissipative quantum chain.
We observe the appearance of an additional topological phase around $\beta = 0$ and $\gamma \approx 5.2$, expanding rapidly toward $\beta \to 0.5$ with increasing $\gamma$ [see Fig.~\ref{fig:local_transport}(a)].
We note that the topological phase on the left is governed by the dominant contribution of a single outlier, whereas the topological phase on the right arises from the collective winding of the outlier and the bulk states.
Consequently, the former cannot exist without quantum jumps, while the latter persists in their presence as a remnant of the underlying non-Hermitian topological phase.
We thus find a \emph{quantum-jump-induced topology} (cf.~Fig.~\ref{fig:specLambda}).
As one can see, the topological phases are characterized by winding numbers with opposite sign, which is also supported by both the steady-state localization and the transport properties, as shown below.

Furthermore, we observe that the mean transport after $p=1$ and $p=5$ Floquet periods [see Fig.~\ref{fig:local_transport}(b) and \ref{fig:local_transport}(c)] approximately follows the topological winding number.
Within the topological phase, the direction of transport agrees with the winding number, although the transport is not quantized.
The transport for $p=1$ is symmetric with respect to $\frac{\beta T}{3} = \frac{\pi}{4}$, whereas this symmetry is broken for $p>1$.
This dynamical signal is consistent with the asymmetry of the underlying topological phases.
In particular, the phase transition, at which the transport direction changes, is compatible with $\gamma=-\frac{3}{T}\ln\cos(\frac{2\beta T}{3})$  in the long-time limit [see dashed line in Fig.~\ref{fig:local_transport}(c)].
This expression, up to a factor of 2, coincides with the phase boundary between positive and negative steady-state spin polarizations for the single-site model studied in Appendix \ref{app:single_site_model}.

In Figs.~\ref{fig:local_transport}(d)-\ref{fig:local_transport}(f), we show the stroboscopic time evolutions of an unpolarized fermion placed in the middle of the quantum chain with open boundary conditions for different parameters.
Here, the main plots show the expectation value of the particle number $\expval{n_j(pT)}$ and the insets show the expectation value of the spin polarization $\expval{\sigma_z^j(pT)}$.
One can see that for Figs.~\ref{fig:local_transport}(d) and \ref{fig:local_transport}(f), the fermion moves continuously to the boundary, becomes partially delocalized during its motion, and eventually gets trapped by the boundary.
In contrast, in Fig.~\ref{fig:local_transport}(e), the fermion becomes delocalized along the full chain and is not trapped at the boundary.
This behavior is consistent with the topological phases shown in Fig.~\ref{fig:local_transport}(a) and the steady-state localization shown in Fig.~\ref{fig:local_skin_effect}(a).

\subsection{Steady-state localization}
\label{ssec:steady_state_loc}
\begin{figure}
    \includegraphics[width=0.99\columnwidth]{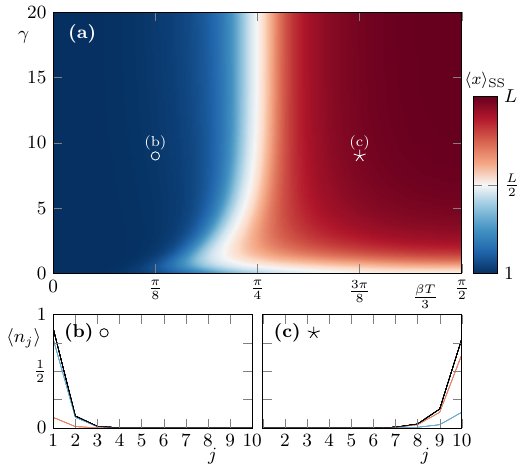}
    \caption{
        Steady-state localization of the Floquet-Liouvillian propagator $\F$.
        (a) The spatial Localization $\expval{x}_\mathrm{SS} = \tr[x \rho_\mathrm{SS}]$ of the steady-state $\rho_\mathrm{SS}$ (the only eigenstate with unit trace) as a function of the system parameters $\beta$ and $\gamma$.
        The parameter sets $(\tfrac{\beta T}{3}, \gamma) = (\tfrac{\pi}{8}, 9)$ and $(\tfrac{3\pi}{8}, 9)$ are highlighted, with their corresponding steady-state distributions shown in panels (b) and (c).
        (b), (c) Spatial density distributions $\tr[n_i \rho_\mathrm{SS}]$ (black) together with the spin-resolved components $\tr[n_{j\sigma}\rho_\mathrm{SS}]$ for $\sigma=\,\uparrow$ (red) and $\sigma=\,\downarrow$ (blue).
        The figure shows that the steady-state localization and the topological phases in Fig.~\ref{fig:local_transport} are qualitatively similar.
        We can thus interpret the localization of the open-boundary steady-state as a signature of the topological properties in the periodic system.
    }
    \label{fig:local_skin_effect}
\end{figure}

We now investigate the localization of the steady-state of the Floquet operator $\F$ with open boundaries.
Figure \ref{fig:local_skin_effect}(a) shows the localization $\tr[x \rho_\mathrm{SS}]$ and Figs.~\ref{fig:local_skin_effect}(b)-\ref{fig:local_skin_effect}(c) show two specific parameter sets of $\beta$ and $\gamma$.
We find that the localization looks qualitatively very similar to the transferred charge after many Floquet steps.
The nose that can be seen near the regime $\gamma \approx 1, \frac{\beta T}{3} \approx \frac{\pi}{6}$ can be traced back to a finite-size effect.
In the thermodynamic limit $L \to \infty$, the white line, where the steady-state is not localized at one side, extends to $\beta = 0, \gamma = 0$ without bending back.

\section{Toward Experimental realization}
\label{sec:experimental_realization}
In this section, we take a look at an experimentally more feasible version of the model.
To this end, we include dephasing terms described by local Lindblad operators $L_j^z = \psi_j^\dagger \sigma_z \psi_j$ with dissipation rate $\gamma_z(t)$.
Furthermore, we consider dissipation at all times $t$, using the following protocol:
\begin{align}
    \gamma_{-/z}(t) & = \begin{cases}
                            \gamma_{-/z}  & \text{if } t \in [0, \frac{T}{3}), \\
                            r\gamma_{-/z} & \text{if } t \in [\frac{T}{3}, T).
                        \end{cases}
\end{align}
Here we use $\gamma_-$ to denote the dissipation rate for jump operators in Eq~\eqref{eqn:Ljs}.
We also include a prefactor $r$, which can be experimentally achieved by either engineering the actual dissipation rates or, practically simpler, staying longer within the waiting phase.

\begin{figure}
    \includegraphics[width=0.99\columnwidth]{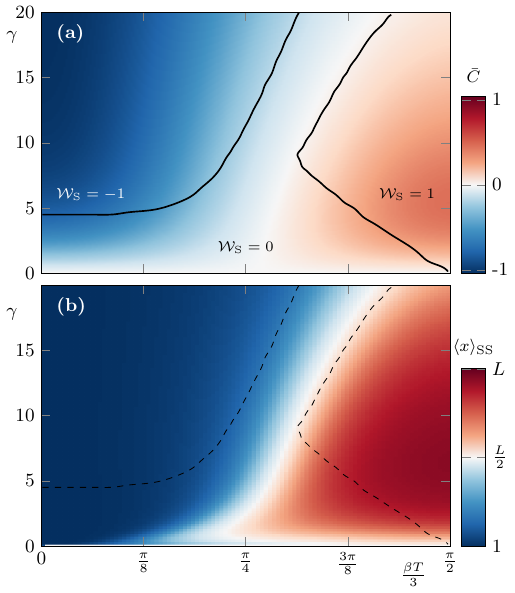}
    \caption{
        Mean transferred charge, steady-state localization, and topological winding number for a realistic model with both spontaneous decay and dephasing at all times.
        Parameters are $\gamma_- = \frac{\gamma_z}{10} = \gamma$ and $r = \frac{1}{10}$.
        (a) Transferred charge $\bar C(1)$ for different $\beta$ and $\gamma$.
        (b) Localization $\expval{x}_\mathrm{SS} = \tr[x \rho_\mathrm{SS}]$ of the steady-state $\rho_\mathrm{SS}$  for different $\beta$ and $\gamma$.
        The topological phase boundaries of the winding number $W$ are indicated by solid lines in panel (a) and dashed lines in panel (b).
        The figure shows that the inclusion of dephasing and the presence of dissipation at all times also allow for the existence of nontrivial topological phases, and that the winding number still qualitatively explains the transport and localization properties of the system.
    }
    \label{fig:experimental_realization}
\end{figure}

In Figure~\ref{fig:experimental_realization}, we show the results for the winding number, the transferred charge $\bar C$, and the steady-state localization.
One can see that the winding number roughly explains the transferred charge and the steady-state localization.

As a concrete experimental platform, one could consider ultracold atoms in optical lattices, where the required Hamiltonian dynamics can be engineered using laser-assisted tunneling and spin-dependent potentials, as it has already been demonstrated in Refs.~\cite{Budich2015,Mancini2015}.
This platform also allows for the implementation of controlled dissipation, for instance, through localized light-induced losses or by coupling to an engineered reservoir.

Note that many-body interactions, which are not included in our single-particle model, are expected to give rise to additional effects, e.g., reflection of particles at each other or joint transport of multiple particles, which can further enrich the topological structure and transport properties of the system.
However, the single-particle sector is still expected to capture the essential topological properties and transport phenomena as long as the interactions between the particles are weak and the particle density is low.

\section{Concluding remarks}
In this work, we have introduced a first generalization of Floquet non-Hermitian topology to open quantum systems.
Our framework builds on the spectral structure of the Liouvillian superoperator and its relation to transport in dissipative Floquet systems.
Specifically, we defined a topological winding number -- extending the construction of Ref.~\cite{Hockendorf2020} -- that arises from the product structure of the no-jump Liouvillian and quantifies the $k$-dependent winding of eigenvalues around the origin of the complex plane.
The inclusion of quantum jumps produces a noticeable shift in a small number of eigenvalues.
We identify the system’s topological winding number with that of these outlier modes, defined either (1) as the dominant spectral contribution or (2) as the modes that wind in the same direction as the bulk without dominating the long-time dynamics.
This leads to a richer topological phase diagram than that of the effective non-Hermitian approximation.

While our construction provides a consistent description of topology in dissipative dynamics, its formulation may be refined further.
In particular, further elaborating on the separation between population and coherence, the weighting function $f_\mathrm{pop}$ quantifying the extent to which each mode contributes to transport-relevant population versus nontransporting coherence could be included into a refined topological invariant.
Such an approach could further strengthen the physical implications of topology in open quantum systems.
Finally, we remark that the topology of the Floquet Liouvillian has implications beyond time-dependent open Floquet systems.
In particular, similar topological structures can arise in the quantum channels associated with discrete quantum feedback control~\cite{Nakagawa2025}.
Although the continuous-time Lindblad master equation and discrete quantum channels differ in their physical interpretation, their topological properties may be unified within a broader framework for topology in open quantum systems.
Another complementary approach to defining a Liouvillian winding number has also been reported in the context of full counting statistics, where an effective momentum variable emerges \cite{Pavlov2025}.
While there are mathematical similarities between their approach and ours, the underlying physical setting and the interpretation of the topological invariant are quite different.

\section*{Acknowledgments}
We would like to thank Carl Lehmann for discussions.
We acknowledge financial support from the German Research Foundation (DFG) through the Collaborative Research Centre SFB 1143 (Project No.~247310070), the Cluster of Excellence ctd.qmat (Project No.~390858490), and the DFG Project No.~459864239.
Our numerical calculations were performed on resources at the TU Dresden Center for Information Services and High Performance Computing (ZIH).

\section*{Data availability}
The data that support the findings of this article are openly available \cite{Zenodo2026}.

\appendix

\section{Vectorization of the translation invariant, single-particle space}
\label{app:vectorization}
In this Appendix, we give a more detailed version of the vectorization scheme.
For the quantum system, we chose $\ket{m,s} = \ket{m} \otimes \ket{s}$ as the basis states, where $m$ denotes the site and $s$ the spin configuration.
The density matrix can then be written as
\begin{align}
    \rho = \sum_{m,m'}\sum_{s,s'} \rho_{(ms),\,(m's')}\ket{m,s}\bra{m',s'}.
\end{align}
We vectorize the density matrix by defining the basis states $\ket{m,s} \otimes (\bra{m',s'})^T$ and then (using the separation of spatial and spin components) changing the order of the parameters to
\begin{align}
    \dket{m,m',s,s'} = \ket{m} \otimes \bra{m'}^T \otimes \ket{s} \otimes \bra{s'}^T.
\end{align}
The procedure maps the density matrix to
\begin{align}
    \rho \to \dket{\rho} = \sum_{m,m'}\sum_{s,s'} \rho_{(ms),\,(m's')} \dket{m,m',s,s'}.
\end{align}

The Lindblad master equation [Eq.~\eqref{eq:Lindblad}] is acting onto the density matrix from the left and the right.
Using the non-Hermitian Hamiltonian [see Eq.~\eqref{eq:NH_Hamiltonian}], it can be written as
\begin{align}
    \dot\rho = -i (H_\mathrm{NH}\rho - \rho H_\mathrm{NH}^\dagger) + \sum_\kappa \gamma_\kappa L_\kappa \rho L_\kappa^\dagger
\end{align}
where the NH Hamiltonian is acting \emph{either} from the left or from the right, and it can be written generically as
\begin{align}
    H_\mathrm{NH} & = \sum_{m,m'} \sum_{s,s'} h_{(ms),\,(m's')} \ket{m,s}\bra{m',s'}.
\end{align}
Assuming translation invariance,
\begin{align}
    \comm{H_\mathrm{NH}}{T_1} = 0, \quad T_1 = \sum_m \ket{m+1}\bra{m} \otimes \mathds{1}_2,
\end{align}
it follows that the matrix elements depend only on the distance $\delta = m-m'$:
\begin{align}
    h_{(ms),\,(m's')} = h^{m-m'}_{ss'} \equiv h^\delta_{s,s'}.
\end{align}
Therefore, the Hamiltonian can be written as a sum of tensor products of spatial and spin operators:
\begin{align}
    H_\mathrm{NH} = \sum_{\delta} R_\delta \otimes S_\delta,
\end{align}
where the spatial component is
\begin{align}
    R_\delta = \sum_m \ket{m+\delta}\bra{m},
\end{align}
and the spin component is
\begin{align}
    S_\delta = \sum_{s,s'} h^\delta_{ss'} \ket{s}\bra{s'}.
\end{align}

Optionally, for spin $\frac{1}{2}$, one can further expand $S_\delta$ in the Pauli basis:
\begin{align}
    S_\delta = c_\delta \, \mathds{1} + \vec{v}_\delta \cdot \vec{\sigma}.
\end{align}

The vectorization of the von-Neumann-like part of the Lindblad master equation then reads
\begin{align}
     & \quad H_\mathrm{NH} \ket{m,s}\bra{m',s'} - \ket{m,s}\bra{m',s'} H_\mathrm{NH}^\dagger                                \\
     & \to \left(H_\mathrm{NH} \otimes \mathds{1} - \mathds{1} \otimes H_\mathrm{NH}^*\right) \ket{m,s}\otimes\bra{m',s'}^T \\
     & \to \sum_\delta (T_\delta \otimes \mathds{1}_L \otimes S_\delta \otimes \mathds{1}_2 \nonumber                       \\
     & \qquad- \mathds{1}_L \otimes T_\delta \otimes \mathds{1}_2 \otimes S_\delta^*) \dket{m,m',s,s'}
\end{align}

Additionally, we also have the so-called quantum jump terms $L_\kappa \rho L_\kappa^\dagger$, the ($\gamma_\kappa$-weighted) sum of which is translation invariant by translation with a single site ($T_1 \mathcal J[\rho] T_1^\dagger = \mathcal J[T_1 \rho T_1^\dagger]$, with $\mathcal J[\rho] = \sum_\kappa \gamma_\kappa L_\kappa \rho L_\kappa^\dagger$) as well.
Here, we only focus on dissipative processes that locally flip the spin at a single site, e.g.,
\begin{align}
    L^\kappa_m = \ket{m}\bra{m} \otimes S_\kappa
\end{align}
The total dissipative contribution to the dynamics is then obtained by applying this process at every site:
\begin{align}
    \mathcal J[\rho] = \sum_\kappa {\gamma_\kappa} \sum_m L_m^\kappa \rho \, (L_m^\kappa)^\dagger.
\end{align}
which ensures translation invariance.
The quantum jump terms can then be written as
\begin{align}
    \mathcal J[\ket{m,s}\bra{m',s'}] \to \sum_\kappa {\gamma_\kappa} P \otimes S_\kappa \otimes S_\kappa^* \dket{m,m',s,s'}
\end{align}
with $P = \sum_m \ket{m}\bra{m}\otimes\ket{m}\bra{m}$ is a $L^2 \times L^2$ matrix.
In the case of spontaneous decay, we represent $S_- \to \sigma_-$, and for dephasing $S_z \to \sigma_z$.
Note that the spin operators $S_\delta$ appearing in the NH Hamiltonian and $S_\kappa$ in the jump operators are neither identical nor independent, but must be chosen consistently such that the Lindblad master equation is satisfied.

At this point, we can write the vectorized Liouvillians in real-space representation as follows:
\begin{align}
    \L_1           & = (-i\beta) \left[\mathds{1}_L \otimes \mathds{1}_L \otimes (\sigma_x \otimes \sigma_0 - \sigma_0 \otimes \sigma_x)\right]                         \\
    \L_2           & = (i\alpha) [T_{-1} \otimes \mathds{1}_L \otimes \sigma_+ \otimes \sigma_0 + T_1 \otimes \mathds{1}_L \otimes \sigma_- \otimes \sigma_0 \nonumber  \\
                   & \quad- \mathds{1} \otimes T_1 \otimes \sigma_0 \otimes \sigma_- - \mathds{1} \otimes T_{-1} \otimes \sigma_0 \otimes \sigma_+]                     \\
    \L^\mathrm{nj} & = -\frac{\gamma}{2} \left[\mathds{1}_L \otimes \mathds{1}_L \otimes (\sigma_+\sigma_- \otimes \sigma_0 + \sigma_0 \otimes \sigma_+\sigma_-)\right] \\
    \L^\mathrm{j}  & = \gamma [P \otimes \sigma_- \otimes \sigma_-]
\end{align}

Now we perform another basis transformation $\dket{m,m',s,s'} \to \dket{m; \Delta,s,s'}$, with $\Delta = m'-m$.
This transformation allows us to write $P = \mathds{1}_L \otimes \delta_{\Delta=0}$.
The Liouvillians can then be written as
\begin{align}
    \L_1           & = (-i\beta) \left[\mathds{1}_L \otimes \mathds{1}_L \otimes (\sigma_x \otimes \sigma_0 - \sigma_0 \otimes \sigma_x)\right]                         \\
    \L_2           & = (i\alpha) [T_{-1} \otimes T_1 \otimes \sigma_+ \otimes \sigma_0 + T_1 \otimes T_{-1} \otimes \sigma_- \otimes \sigma_0 \nonumber                 \\
                   & \quad- \mathds{1} \otimes T_1 \otimes \sigma_0 \otimes \sigma_- - \mathds{1} \otimes T_{-1} \otimes \sigma_0 \otimes \sigma_+]                     \\
    \L^\mathrm{nj} & = -\frac{\gamma}{2} \left[\mathds{1}_L \otimes \mathds{1}_L \otimes (\sigma_+\sigma_- \otimes \sigma_0 + \sigma_0 \otimes \sigma_+\sigma_-)\right] \\
    \L^\mathrm{j}  & = \gamma [\mathds{1}_L \otimes \delta_{\Delta = 0} \otimes \sigma_- \otimes \sigma_-]
\end{align}
We thus can perform a Fourier transform with respect to $m$:
\begin{align}
    \dket{k;\Delta,s,s'} = \frac{1}{\sqrt{L}}\sum_m e^{-ikm} \dket{m;\Delta,s,s'}
\end{align}
which block diagonalizes the Liouvillians, resulting in
\begin{align}
    \L_1(k)           & = (-i\beta) \left[\mathds{1}_L \otimes (\sigma_x \otimes \sigma_0 - \sigma_0 \otimes \sigma_x)\right]                         \\
    \L_2(k)           & = (i\alpha) [e^{-ik} T_1 \otimes \sigma_+ \otimes \sigma_0 + e^{ik} T_{-1} \otimes \sigma_- \otimes \sigma_0 \nonumber        \\
                      & \quad- T_1 \otimes \sigma_0 \otimes \sigma_- - T_{-1} \otimes \sigma_0 \otimes \sigma_+]                                      \\
    \L^\mathrm{nj}(k) & = -\frac{\gamma}{2} \left[\mathds{1}_L \otimes (\sigma_+\sigma_- \otimes \sigma_0 + \sigma_0 \otimes \sigma_+\sigma_-)\right] \\
    \L^\mathrm{j}(k)  & = \gamma [\delta_{\Delta = 0} \otimes \sigma_- \otimes \sigma_-]
\end{align}
This form of the Liouvillians is used to calculate the winding numbers for the open quantum system.

In the non-Hermitian approximation, where quantum jumps are neglected, we can further apply a similar Fourier transform to the $\Delta$ index.
\begin{align}
    \L_1(q, k)           & = (-i\beta) [\sigma_x \oplus (-\sigma_x)]                                    \\
    \L_2(q, k)           & = (i\alpha) \bigl[(e^{-i(k - q)} \sigma_+ + e^{i(k - q)} \sigma_-) \nonumber \\
                         & \qquad\oplus \left(- e^{iq}\sigma_- - e^{-iq} \sigma_+\right)\bigr]          \\
    \L^\mathrm{nj}(q, k) & = -\frac{\gamma}{2} [\sigma_+\sigma_- \oplus \sigma_+\sigma_-]
\end{align}
where we introduced the Kronecker sum $A \oplus B = A\otimes \mathds{1} + \mathds{1} \otimes B$.
Using the Bloch Hamiltonians $H_1(k) = \beta \sigma_x$, $H_2(k) = -\alpha[\cos(k) \sigma_x - \sin(k) \sigma_y]$, and $H_\mathrm{NH}(k) = -\frac{i\gamma}{2} \sigma_+\sigma_-$, we can write the vectorized Lindbladians in the form
\begin{align}
    \L_j(q, k) & = (-i H_j(q-k)) \oplus (-i H_j(q))^*.
\end{align}
This simple form is very helpful for calculating the Lindbladian Floquet propagator
\begin{align}
    \F(q, k) & = \prod_j \exp\left(\L_j(q, k) \frac{T}{3}\right) \\
             & = F(q - k) \otimes F(q)^*
\end{align}
where we used the matrix properties $\exp(A \oplus B) = \exp(A) \otimes \exp(B)$, $\exp(A^*) = \exp(A)^*$, and $A^* B^* = (A B)^*$.
The NH Floquet operator is $F(k) = \prod_j \exp(-i H_j(k))$.
It allows us to link the topological behavior of the NH Floquet operator $F(k)$ to the Lindbladian Floquet operator $\F(q, k)$ by fixing a momentum $q=\tilde q$.

\section{Transferred charge in density matrix formalism}
\label{app:transferred_charge}
We define the mean transferred charge over one Floquet period by
\begin{align}
    \bar C = \expval{x - n}_{\F[\rho_n(0)]}
\end{align}
with position operator $x = \sum_{m, s} m \ket{m,s}\bra{m,s}$, Floquet propagator $\F$, and initial state $\rho_n(0) = \frac{1}{2} \sum_r \ket{n,r}\bra{n,r}$.
After vectorizing the density matrix ($\ket{m,s}\bra{m',s'} \to \dket{m,m',s,s'}$), this becomes
\begin{align}
    \bar C = \frac{1}{2} \sum_m \sum_{s,r} (m-n)\dbra{m,m,s,s} \F \dket{n,n,r,r}.
\end{align}
Applying a basis transformation $\dket{m,m',s,s'} \to \dket{m;\Delta=m'-m,s,s'}$ and a Fourier transform $\dket{k;\Delta,s,s'} = \frac{1}{\sqrt{L}}\sum_m e^{-ikm} \dket{m;\Delta,s,s'}$ (with inverse $\dket{m;\Delta,s,s'} = \frac{1}{\sqrt{L}}\sum_m e^{ikm} \dket{k;\Delta,s,s'}$ diagonalizes the translation invariant Floquet propagator with respect to $k$ (see Appendix \ref{app:vectorization} for more details).
We then have
\begin{align}
     & \langle\!\langle m;\Delta,s,s' | k;\Delta',r,r' \rangle\!\rangle \nonumber                            \\
     & \quad= \tr[\ket{\Delta,s'}\bra{m,s} \frac{1}{\sqrt{L}}\sum_{m'} e^{-ikm'} \ket{m',r}\bra{\Delta',r'}] \\
     & \quad= \frac{1}{\sqrt{L}}e^{-ikm} \delta_{\Delta,\Delta'} \delta_{sr} \delta_{s'r'}
\end{align}
which we can use to calculate the mean transferred charge
\begin{align}
    \bar C & = \frac{1}{2}\sum_{s,r} \sum_m (m-n) \dbra{m;0,s,s}  \nonumber                                              \\
           & \quad \left(\sum_k \sum_{\nu_1,\nu_2}\F_{\nu_1,\nu_2}(k)\dket{k;\nu_1} \dbra{k;\nu_2}\right) \dket{n;0,r,r} \\
           & = \frac{1}{2L}\sum_{\substack{s,r                                                                           \\ m,k}} (m-n) e^{-ik(m-n)} \F_{(0ss),\,(0rr)}(k) \\
           & = \frac{1}{2L}\sum_{\substack{s,r                                                                           \\ m,k}} \left[i\pdv{k} e^{-ik(m-n)}\right] \F_{(0ss),\,(0rr)}(k).
\end{align}
Here, we used the multi-indices $\nu_i = (\Delta_i,s_i,r_i)$.
In the continuum limit, this becomes
\begin{align}
     & = \frac{1}{4\pi}\sum_{s,r} \int_0^{2\pi} \sum_m \left[i\pdv{k} e^{-ik(m-n)}\right] \F_{(0ss),\,(0rr)}(k) \dd k                                  \\
     & = \frac{(-i)}{4\pi}\sum_{s,r} \int_0^{2\pi} \underbrace{\sum_m  e^{-ik(m-n)}}_{2\pi \delta(k)} \left[\pdv{k}\F_{(0ss),\,(0rr)}(k) \right] \dd k \\
     & = \frac{(-i)}{2}\sum_{s,r} \pdv{k}\F_{(0ss),\,(0rr)}(k)\Big|_{k=0}
\end{align}

Without the quantum jump terms, we apply another Fourier transform to the $\Delta$ indices and write
\begin{align}
     & \F_{(\Delta ss'),\,(\Delta'rr')}(k) \nonumber                                              \\
     & \quad= \frac{1}{2\pi} \int_0^{2\pi} e^{-iq(\Delta-\Delta')} \F_{(ss'),\,(rr')}(k, q) \dd q
\end{align}
We further find that $\F(k, q) = F(q-k) \otimes F^*(q)$ (see Appendix \ref{app:vectorization}), which reads with indices inserted $\F_{(ss'),\,(rr')}(k, q) = F_{sr}(q-k) F_{s'r'}^*(q)$.
Inserting this into the mean transferred charge, we find
\begin{align}
    \bar C^\mathrm{nj} & = \frac{(-i)}{4\pi} \int_0^{2\pi} \sum_{s,r} \pdv{k} F_{sr}(q-k) F^*_{sr}(q)\Big|_{k=0} \dd q       \\
                       & = \frac{(-i)}{4\pi} \int_0^{2\pi} \sum_{s,r} \pdv{k} F_{sr}(q-k) F^\dagger_{rs}(q)\Big|_{k=0} \dd q \\
                       & = \frac{(-i)}{4\pi} \int_0^{2\pi} \tr_S\left[\pdv{k} F(q-k) F^\dagger(q)\right]\Big|_{k=0} \dd q    \\
                       & = \frac{(-i)}{4\pi} \int_0^{2\pi} \tr_S\left[F^\dagger(q) \pdv{q} F(q)\right] \dd q
\end{align}
where $\tr_S A = \sum_\sigma A_{\sigma\sigma}$ sums over the spin degrees of freedom of the Fourier transform.

\section{Single-site open quantum system}
\label{app:single_site_model}
In this appendix, we investigate a related Floquet open quantum system, which simply includes one spin degree of freedom.
In other words, we turn off the couplings between different unit cells.
The time evolution of the single-site density matrix $\rho(t)$ is still controlled by the time-dependent Lindblad master equation in Eq.~\eqref{eq:Lindblad}.
With $\sigma_{x,y,z}$ representing the Pauli operators for the spin degree of freedom, the full Liouvillian dynamics in one period $T$ is given by
\begin{align}
    H_\mathrm{S}(t) & = \begin{cases}
                            0              & \text{if } t \in [0, \frac{T}{3}),             \\
                            \beta \sigma_x & \text{if } t \in [\frac{T}{3},  \frac{2T}{3}), \\
                            0              & \text{if } t \in [\frac{2T}{3}, T).
                        \end{cases} \\
    \gamma(t)       & = \begin{cases}
                            \gamma & \text{if } t \in [0, \frac{T}{3}), \\
                            0      & \text{if } t \in [\frac{T}{3}, T).
                        \end{cases}
\end{align}
The jump operator is given by $L=\sigma_-$.
Here, the third time interval without any dynamics is introduced to match the drive protocol for the one-dimensional system studied in the main text.
The full dynamics in one period is generated by the Floquet-Liouvillian propagator $\F=\exp(\L_3 T/3)\exp(\L_2 T/3)\exp(\L_1 T/3)$.
After the vectorization of the density matrix, the Floquet-Lindbladian superoperator has the following matrix form:
\begin{align}
    \F & =\exp\left(-\frac{i\beta T}{3}\left(\sigma_x \oplus (-\sigma_x)\right)\right) \nonumber                                       \\
       & \quad\times\exp\left(\frac{\gamma T}{3}(\sigma^-\otimes \sigma^- - \frac{1}{2}\sigma^+\sigma^-\oplus \sigma^+\sigma^-)\right)
\end{align}

\begin{figure}
    \includegraphics[width=0.99\columnwidth]{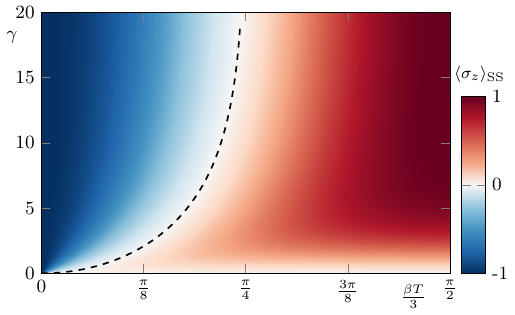}
    \caption{
        Expectation value of the steady-state spin polarization $\expval{\sigma_z} = \tr[\sigma_z \rho_\mathrm{SS}]$ of the single-site model.
        The dashed line displays the phase boundary where a sign change appears [cf.~Eq.~\eqref{eq:single_site_polarization}].
    }
    \label{afig:single}
\end{figure}

Defining $\theta=\beta T/3$, $G=\exp(\gamma T/6)$, and $D=G^2 - 2G \cos^2\theta + 1$, we find that the steady-state density matrix $\rho$, which satisfies $\F[\rho]=\rho$, has the following expression:
\begin{equation}
    \rho=\frac{1}{D}\begin{pmatrix}
        G^2 \sin^2\theta
         &
        \frac{i}{2} G(G - 1) \sin2\theta
        \\
        -\frac{i}{2}G(G - 1) \sin2\theta
         &
        D - G^2 \sin^2\theta
    \end{pmatrix}.
\end{equation}
Based on this expression, we find that the steady-state spin polarization is given by
\begin{equation}
    P=\tr[\rho\sigma_z]=\frac{\cos^2\theta(2G-2G^2)-1+G^2}{D}
\end{equation}
Then, we can identify a phase boundary where $P$ changes its sign.
By solving $P=0$, we obtain $G\cos(2\theta)=1$, which leads to a phase boundary shown in Fig.~\ref{afig:single}:
\begin{equation}
    \label{eq:single_site_polarization}
    \gamma=-\frac{6}{T}\ln\cos\left(\frac{2\beta T}{3}\right).
\end{equation}

After turning on the hopping between different unit cells, the different spin components prefer to move in the opposite direction, and a nonzero polarization is expected to induce chiral transport.
Nevertheless, the interplay between intercell and intracell couplings will quantitatively change the phase boundary, as shown in Fig.~\ref{fig:local_transport}(c).


\bibliography{literature.bib}

@article{Ashida2020,
  title     = {Non-Hermitian Physics},
  author    = {Ashida, Yuto and Gong, Zongping and Ueda, Masahito},
  year      = 2020,
  month     = jul,
  journal   = {Adv. Phys.},
  volume    = {69},
  number    = {3},
  pages     = {249--435},
  publisher = {Informa UK Limited},
  issn      = {1460-6976},
  doi       = {10.1080/00018732.2021.1876991}
}

@article{Benaych-Georges2011,
  title     = {The Eigenvalues and Eigenvectors of Finite, Low Rank Perturbations of Large Random Matrices},
  author    = {{Benaych-Georges}, Florent and Nadakuditi, Raj Rao},
  year      = 2011,
  month     = may,
  journal   = {Advances in Mathematics},
  volume    = {227},
  number    = {1},
  pages     = {494--521},
  issn      = {00018708},
  doi       = {10.1016/j.aim.2011.02.007},
  urldate   = {2025-11-13},
  copyright = {https://www.elsevier.com/tdm/userlicense/1.0/},
  langid    = {english}
}

@article{Bergholtz2021,
  title     = {Exceptional Topology of Non-{{Hermitian}} Systems},
  author    = {Bergholtz, Emil J. and Budich, Jan Carl and Kunst, Flore K.},
  year      = 2021,
  month     = feb,
  journal   = {Rev. Mod. Phys.},
  volume    = {93},
  number    = {1},
  pages     = {015005},
  publisher = {American Physical Society},
  doi       = {10.1103/RevModPhys.93.015005}
}

@misc{Bordenave2024,
  title         = {Outliers of Perturbations of Banded {{Toeplitz}} Matrices},
  author        = {Bordenave, Charles and Chapon, Fran{\c c}ois and Capitaine, Mireille},
  year          = 2024,
  month         = oct,
  number        = {arXiv:2410.16439},
  eprint        = {2410.16439},
  primaryclass  = {math},
  publisher     = {arXiv},
  doi           = {10.48550/arXiv.2410.16439},
  urldate       = {2025-11-13},
  archiveprefix = {arXiv}
}

@book{Breuer2009,
  title     = {The Theory of Open Quantum Systems},
  author    = {Breuer, Heinz-Peter and Petruccione, Francesco},
  year      = 2009,
  edition   = {1. publ. in paperback, [Nachdr.]},
  publisher = {Clarendon Press},
  address   = {Oxford [u.a.]},
  isbn      = {978-0-19-852063-4}
}

@article{Budich2015,
  title     = {Synthetic helical liquids with ultracold atoms in optical lattices},
  author    = {Budich, J. C. and Laflamme, C. and Tschirsich, F. and Montangero, S. and Zoller, P.},
  journal   = {Phys. Rev. B},
  volume    = {92},
  issue     = {24},
  pages     = {245121},
  numpages  = {9},
  year      = {2015},
  month     = {Dec},
  publisher = {American Physical Society},
  doi       = {10.1103/PhysRevB.92.245121},
  url       = {https://link.aps.org/doi/10.1103/PhysRevB.92.245121}
}

@article{Budich2016,
  title     = {Dynamical Topological Order Parameters Far from Equilibrium},
  author    = {Budich, Jan Carl and Heyl, Markus},
  year      = 2016,
  month     = feb,
  journal   = {Phys. Rev. B},
  volume    = {93},
  number    = {8},
  pages     = {085416},
  publisher = {American Physical Society},
  doi       = {10.1103/PhysRevB.93.085416}
}

@article{Budich2017,
  title     = {Helical Floquet Channels in {{1D}} Lattices},
  author    = {Budich, Jan Carl and Hu, Ying and Zoller, Peter},
  year      = 2017,
  month     = mar,
  journal   = {Phys. Rev. Lett.},
  volume    = {118},
  number    = {10},
  pages     = {105302},
  publisher = {American Physical Society (APS)},
  issn      = {1079-7114},
  doi       = {10.1103/physrevlett.118.105302}
}

@article{Budich2020,
  title     = {Non-Hermitian Topological Sensors},
  author    = {Budich, Jan Carl and Bergholtz, Emil J.},
  year      = 2020,
  month     = oct,
  journal   = {Phys. Rev. Lett.},
  volume    = {125},
  number    = {18},
  pages     = {180403},
  publisher = {American Physical Society},
  doi       = {10.1103/PhysRevLett.125.180403}
}

@article{Cayssol2013,
  title    = {Floquet Topological Insulators},
  author   = {Cayssol, J{\'e}r{\^o}me and D{\'o}ra, Bal{\'a}zs and Simon, Ferenc and Moessner, Roderich},
  year     = 2013,
  journal  = {Phys. Status Solidi RRL -- Rapid Res. Lett.},
  volume   = {7},
  number   = {1-2},
  pages    = {101--108},
  doi      = {10.1002/pssr.201206451}
}

@article{Chaduteau2025,
  title     = {Lindbladian versus Postselected non-Hermitian Topology},
  author    = {Chaduteau, Alexandre and Lee, Derek K. K. and Schindler, Frank},
  journal   = {Phys. Rev. Lett.},
  volume    = {136},
  issue     = {1},
  pages     = {016603},
  numpages  = {6},
  year      = {2026},
  month     = {Jan},
  publisher = {American Physical Society},
  doi       = {10.1103/ljvt-w6hw},
  url       = {https://link.aps.org/doi/10.1103/ljvt-w6hw}
}

@article{Chen2025,
  title     = {Engineering Nonequilibrium Steady States through Floquet Liouvillians},
  author    = {Chen, Weijian and Abbasi, Maryam and Erdamar, Serra and Muldoon, Jacob and Joglekar, Yogesh N. and Murch, Kater W.},
  year      = 2025,
  month     = mar,
  journal   = {Phys. Rev. Lett.},
  volume    = {134},
  number    = {9},
  pages     = {090402},
  publisher = {American Physical Society},
  doi       = {10.1103/PhysRevLett.134.090402}
}

@article{Dai2016,
  title     = {Floquet Theorem with Open Systems and Its Applications},
  author    = {Dai, C. M. and Shi, Z. C. and Yi, X. X.},
  year      = 2016,
  month     = mar,
  journal   = {Phys. Rev. A},
  volume    = {93},
  number    = {3},
  pages     = {032121},
  publisher = {American Physical Society (APS)},
  issn      = {2469-9934},
  doi       = {10.1103/physreva.93.032121}
}

@article{Daley2014,
  title     = {Quantum Trajectories and Open Many-Body Quantum Systems},
  author    = {Daley, Andrew J.},
  year      = 2014,
  month     = mar,
  journal   = {Adv. Phys.},
  volume    = {63},
  number    = {2},
  pages     = {77--149},
  publisher = {Informa UK Limited},
  issn      = {1460-6976},
  doi       = {10.1080/00018732.2014.933502}
}

@article{Dash2024,
  title     = {Floquet Exceptional Topological Insulator},
  author    = {Dash, Gaurab Kumar and Bid, Subhajyoti and Thakurathi, Manisha},
  year      = 2024,
  month     = jan,
  journal   = {Phys. Rev. B},
  volume    = {109},
  number    = {3},
  pages     = {035418},
  publisher = {American Physical Society},
  doi       = {10.1103/PhysRevB.109.035418}
}

@article{Eckardt2017,
  title     = {Colloquium: {{Atomic}} Quantum Gases in Periodically Driven Optical Lattices},
  author    = {Eckardt, Andr{\'e}},
  year      = 2017,
  month     = mar,
  journal   = {Rev. Mod. Phys.},
  volume    = {89},
  number    = {1},
  pages     = {011004},
  publisher = {American Physical Society},
  doi       = {10.1103/RevModPhys.89.011004}
}

@article{Flaschner2018,
  title   = {Observation of Dynamical Vortices after Quenches in a System with Topology},
  author  = {Fl{\"a}schner, N. and Vogel, D. and Tarnowski, M. and Rem, B. S. and L{\"u}hmann, D.-S. and Heyl, M. and Budich, J. C. and Mathey, L. and Sengstock, K. and Weitenberg, C.},
  year    = 2018,
  journal = {Nat. Phys.},
  volume  = {14},
  number  = {3},
  pages   = {265--268},
  doi     = {10.1038/s41567-017-0013-8}
}

@book{Gardiner2010,
  title     = {Quantum Noise: A Handbook of Markovian and Non-{{Markovian}} Quantum Stochastic Methods with Applications to Quantum Optics},
  author    = {Gardiner, Crispin W. and Zoller, Peter},
  year      = 2010,
  series    = {Springer Complexity},
  edition   = {3. ed., [Nachdr.]},
  publisher = {Springer},
  address   = {Berlin},
  isbn      = {3-540-22301-0}
}

@article{Geng2023,
  title     = {Nonreciprocal Charge and Spin Transport Induced by Non-{{Hermitian}} Skin Effect in Mesoscopic Heterojunctions},
  author    = {Geng, H. and Wei, J. Y. and Zou, M. H. and Sheng, L. and Chen, Wei and Xing, D. Y.},
  year      = 2023,
  month     = jan,
  journal   = {Phys. Rev. B},
  volume    = {107},
  number    = {3},
  pages     = {035306},
  publisher = {American Physical Society},
  doi       = {10.1103/PhysRevB.107.035306}
}

@article{Gomez-Leon2013,
  title     = {Floquet-Bloch Theory and Topology in Periodically Driven Lattices},
  author    = {{G{\'o}mez-Le{\'o}n}, A. and Platero, G.},
  year      = 2013,
  month     = may,
  journal   = {Phys. Rev. Lett.},
  volume    = {110},
  number    = {20},
  pages     = {200403},
  publisher = {American Physical Society},
  doi       = {10.1103/PhysRevLett.110.200403}
}

@article{Gong2015,
  title     = {Stabilizing Non-{{Hermitian}} Systems by Periodic Driving},
  author    = {Gong, Jiangbin and Wang, Qing-hai},
  year      = 2015,
  journal   = {Phys. Rev. A},
  volume    = {91},
  number    = {4},
  pages     = {042135},
  publisher = {APS}
}

@article{Gong2018,
  title     = {Topological Phases of Non-Hermitian Systems},
  author    = {Gong, Zongping and Ashida, Yuto and Kawabata, Kohei and Takasan, Kazuaki and Higashikawa, Sho and Ueda, Masahito},
  year      = 2018,
  month     = sep,
  journal   = {Phys. Rev. X},
  volume    = {8},
  number    = {3},
  pages     = {031079},
  publisher = {American Physical Society},
  doi       = {10.1103/PhysRevX.8.031079}
}

@article{Gorini1976,
  title     = {Completely Positive Dynamical Semigroups of N-Level Systems},
  author    = {Gorini, Vittorio and Kossakowski, Andrzej and Sudarshan, E. C. G.},
  year      = 1976,
  month     = may,
  journal   = {J. Math. Phys.},
  volume    = {17},
  number    = {5},
  pages     = {821--825},
  publisher = {AIP Publishing},
  issn      = {1089-7658},
  doi       = {10.1063/1.522979}
}

@article{Griffiths1993,
  title     = {Consistent Interpretation of Quantum Mechanics Using Quantum Trajectories},
  author    = {Griffiths, Robert B.},
  year      = 1993,
  month     = apr,
  journal   = {Phys. Rev. Lett.},
  volume    = {70},
  number    = {15},
  pages     = {2201--2204},
  publisher = {American Physical Society (APS)},
  issn      = {0031-9007},
  doi       = {10.1103/physrevlett.70.2201}
}

@article{Hockendorf2020,
  title     = {Topological Origin of Quantized Transport in Non-{{Hermitian}} Floquet Chains},
  author    = {H{\"o}ckendorf, Bastian and Alvermann, Andreas and Fehske, Holger},
  year      = 2020,
  month     = may,
  journal   = {Phys. Rev. Res.},
  volume    = {2},
  number    = {2},
  pages     = {023235},
  publisher = {American Physical Society (APS)},
  issn      = {2643-1564},
  doi       = {10.1103/physrevresearch.2.023235}
}

@book{Horn1994,
  title     = {Topics in Matrix Analysis},
  author    = {Horn, Roger A. and Johnson, Charles R.},
  year      = 1994,
  publisher = {Cambridge University Press},
  address   = {Cambridge; New York},
  isbn      = {0-521-46713-6}
}

@article{Jiang2011,
  title     = {Majorana Fermions in Equilibrium and in Driven Cold-Atom Quantum Wires},
  author    = {Jiang, Liang and Kitagawa, Takuya and Alicea, Jason and Akhmerov, A. R. and Pekker, David and Refael, Gil and Cirac, J. Ignacio and Demler, Eugene and Lukin, Mikhail D. and Zoller, Peter},
  year      = 2011,
  month     = jun,
  journal   = {Phys. Rev. Lett.},
  volume    = {106},
  number    = {22},
  pages     = {220402},
  publisher = {American Physical Society},
  doi       = {10.1103/PhysRevLett.106.220402}
}

@article{Junk2020,
  title     = {Floquet Oscillations in Periodically Driven Dirac Systems},
  author    = {Junk, Vanessa and Reck, Phillipp and Gorini, Cosimo and Richter, Klaus},
  year      = 2020,
  month     = apr,
  journal   = {Phys. Rev. B},
  volume    = {101},
  number    = {13},
  pages     = {134302},
  publisher = {American Physical Society},
  doi       = {10.1103/PhysRevB.101.134302}
}

@article{Kane2005,
  title     = {Quantum Spin Hall Effect in Graphene},
  author    = {Kane, C. L. and Mele, E. J.},
  year      = 2005,
  month     = nov,
  journal   = {Phys. Rev. Lett.},
  volume    = {95},
  number    = {22},
  pages     = {226801},
  publisher = {American Physical Society},
  doi       = {10.1103/PhysRevLett.95.226801}
}

@book{Kato1995,
  title     = {Perturbation {{Theory}} for {{Linear Operators}}},
  author    = {Kato, Tosio},
  year      = 1995,
  series    = {Classics in {{Mathematics}}},
  volume    = {132},
  publisher = {Springer Berlin Heidelberg},
  address   = {Berlin, Heidelberg},
  doi       = {10.1007/978-3-642-66282-9},
  urldate   = {2025-11-13},
  copyright = {http://www.springer.com/tdm},
  isbn      = {978-3-540-58661-6 978-3-642-66282-9}
}

@article{Kawabata2019,
  title     = {Symmetry and Topology in Non-Hermitian Physics},
  author    = {Kawabata, Kohei and Shiozaki, Ken and Ueda, Masahito and Sato, Masatoshi},
  year      = 2019,
  month     = oct,
  journal   = {Phys. Rev. X},
  volume    = {9},
  number    = {4},
  pages     = {041015},
  publisher = {American Physical Society},
  doi       = {10.1103/PhysRevX.9.041015}
}

@article{Kitagawa2010,
  title     = {Topological Characterization of Periodically Driven Quantum Systems},
  author    = {Kitagawa, Takuya and Berg, Erez and Rudner, Mark and Demler, Eugene},
  year      = 2010,
  month     = dec,
  journal   = {Phys. Rev. B},
  volume    = {82},
  number    = {23},
  pages     = {235114},
  publisher = {American Physical Society (APS)},
  issn      = {1550-235X},
  doi       = {10.1103/physrevb.82.235114}
}

@article{Kitagawa2012,
  title     = {Observation of Topologically Protected Bound States in Photonic Quantum Walks},
  author    = {Kitagawa, Takuya and Broome, Matthew A. and Fedrizzi, Alessandro and Rudner, Mark S. and Berg, Erez and Kassal, Ivan and {Aspuru-Guzik}, Al{\'a}n and Demler, Eugene and White, Andrew G.},
  year      = 2012,
  month     = jun,
  journal   = {Nat. Commun.},
  volume    = {3},
  number    = {1},
  publisher = {{Springer Science and Business Media LLC}},
  issn      = {2041-1723},
  doi       = {10.1038/ncomms1872}
}

@article{Koch2022,
  title     = {Quantum Non-{{Hermitian}} Topological Sensors},
  author    = {Koch, Florian and Budich, Jan Carl},
  year      = 2022,
  month     = feb,
  journal   = {Phys. Rev. Res.},
  volume    = {4},
  number    = {1},
  pages     = {013113},
  publisher = {American Physical Society},
  doi       = {10.1103/PhysRevResearch.4.013113}
}

@article{Koch2024,
  title     = {Dissipative Frequency Converter: {{From}} Lindblad Dynamics to Non-{{Hermitian}} Topology},
  author    = {Koch, Florian and Budich, Jan Carl},
  year      = 2024,
  month     = aug,
  journal   = {Phys. Rev. Res.},
  volume    = {6},
  number    = {3},
  pages     = {033124},
  publisher = {American Physical Society},
  doi       = {10.1103/PhysRevResearch.6.033124}
}

@article{Kunst2018,
  title     = {Biorthogonal Bulk-Boundary Correspondence in Non-Hermitian Systems},
  author    = {Kunst, Flore K. and Edvardsson, Elisabet and Budich, Jan Carl and Bergholtz, Emil J.},
  year      = 2018,
  month     = jul,
  journal   = {Phys. Rev. Lett.},
  volume    = {121},
  number    = {2},
  pages     = {026808},
  publisher = {American Physical Society},
  doi       = {10.1103/PhysRevLett.121.026808}
}

@article{Lee2016,
  title     = {Anomalous Edge State in a Non-Hermitian Lattice},
  author    = {Lee, Tony E.},
  year      = 2016,
  month     = apr,
  journal   = {Phys. Rev. Lett.},
  volume    = {116},
  number    = {13},
  pages     = {133903},
  publisher = {American Physical Society},
  doi       = {10.1103/PhysRevLett.116.133903}
}

@article{Lee2019,
  title     = {Anatomy of Skin Modes and Topology in Non-{{Hermitian}} Systems},
  author    = {Lee, Ching Hua and Thomale, Ronny},
  year      = 2019,
  month     = may,
  journal   = {Phys. Rev. B},
  volume    = {99},
  number    = {20},
  pages     = {201103},
  publisher = {American Physical Society},
  doi       = {10.1103/PhysRevB.99.201103}
}

@article{Lin2023,
  title     = {Topological Non-{{Hermitian}} Skin Effect},
  author    = {Lin, Rijia and Tai, Tommy and Li, Linhu and Lee, Ching Hua},
  year      = 2023,
  month     = jul,
  journal   = {Front. Phys.},
  volume    = {18},
  number    = {5},
  publisher = {China Engineering Science Press Co. Ltd.},
  issn      = {2095-0470},
  doi       = {10.1007/s11467-023-1309-z}
}

@article{Lindblad1976,
  title     = {On the Generators of Quantum Dynamical Semigroups},
  author    = {Lindblad, G.},
  year      = 1976,
  month     = jun,
  journal   = {Commun. Math. Phys.},
  volume    = {48},
  number    = {2},
  pages     = {119--130},
  publisher = {{Springer Science and Business Media LLC}},
  issn      = {1432-0916},
  doi       = {10.1007/bf01608499}
}

@article{Lindner2011,
  title     = {Floquet Topological Insulator in Semiconductor Quantum Wells},
  author    = {Lindner, Netanel H. and Refael, Gil and Galitski, Victor},
  year      = 2011,
  month     = mar,
  journal   = {Nat. Phys.},
  volume    = {7},
  number    = {6},
  pages     = {490--495},
  publisher = {{Springer Science and Business Media LLC}},
  issn      = {1745-2481},
  doi       = {10.1038/nphys1926}
}

@article{Liu2022,
  title     = {Symmetry and Topological Classification of Floquet Non-{{Hermitian}} Systems},
  author    = {Liu, Chun-Hui and Hu, Haiping and Chen, Shu},
  year      = 2022,
  month     = jun,
  journal   = {Phys. Rev. B},
  volume    = {105},
  number    = {21},
  pages     = {214305},
  publisher = {American Physical Society},
  doi       = {10.1103/PhysRevB.105.214305}
}

@article{Longhi2017,
  title     = {Floquet Exceptional Points and Chirality in Non-{{Hermitian}} Hamiltonians},
  author    = {Longhi, Stefano},
  year      = 2017,
  journal   = {J. Phys. Math. Theor.},
  volume    = {50},
  number    = {50},
  pages     = {505201},
  publisher = {IOP Publishing}
}

@article{Mancini2015,
  author    = {Mancini, M. and Pagano, G. and Cappellini, G. and Livi, L. and Rider, M. and Catani, J. and Sias, C. and Zoller, P. and Inguscio, M. and Dalmonte, M. and Fallani, L.},
  journal   = {Science},
  title     = {Observation of chiral edge states with neutral fermions in synthetic Hall ribbons},
  year      = {2015},
  issn      = {1095-9203},
  month     = sep,
  number    = {6255},
  pages     = {1510--1513},
  volume    = {349},
  doi       = {10.1126/science.aaa8736},
  publisher = {American Association for the Advancement of Science (AAAS)}
}

@article{Martin2017,
  title     = {Topological Frequency Conversion in Strongly Driven Quantum Systems},
  author    = {Martin, Ivar and Refael, Gil and Halperin, Bertrand},
  year      = 2017,
  month     = oct,
  journal   = {Phys. Rev. X},
  volume    = {7},
  number    = {4},
  pages     = {041008},
  publisher = {American Physical Society},
  doi       = {10.1103/PhysRevX.7.041008}
}

@article{Mehl2011,
  title    = {Eigenvalue Perturbation Theory of Classes of Structured Matrices under Generic Structured Rank One Perturbations},
  author   = {Mehl, Christian and Mehrmann, Volker and Ran, Andr{\'e} C. M. and Rodman, Leiba},
  year     = 2011,
  month    = aug,
  journal  = {Linear Algebra and its Applications},
  series   = {Special {{Issue}}: {{Dedication}} to {{Pete Stewart}} on the Occasion of His 70th Birthday},
  volume   = {435},
  number   = {3},
  pages    = {687--716},
  issn     = {0024-3795},
  doi      = {10.1016/j.laa.2010.07.025},
  urldate  = {2025-11-13}
}

@article{Minganti2020,
  title     = {Hybrid-Liouvillian Formalism Connecting Exceptional Points of Non-{{Hermitian}} Hamiltonians and Liouvillians via Postselection of Quantum Trajectories},
  author    = {Minganti, Fabrizio and Miranowicz, Adam and Chhajlany, Ravindra W. and Arkhipov, Ievgen I. and Nori, Franco},
  year      = 2020,
  month     = jun,
  journal   = {Phys. Rev. A},
  volume    = {101},
  number    = {6},
  pages     = {062112},
  publisher = {American Physical Society},
  doi       = {10.1103/PhysRevA.101.062112}
}

@article{Monkman2026,
  title     = {Limits of the non-Hermitian description of decay models},
  author    = {Monkman, Kyle and Berciu, Mona},
  journal   = {Phys. Rev. A},
  volume    = {113},
  number    = {3},
  pages     = {032213},
  numpages  = {12},
  year      = {2026},
  month     = {3},
  publisher = {American Physical Society},
  doi       = {10.1103/dyjw-1cbl},
  url       = {https://link.aps.org/doi/10.1103/dyjw-1cbl}
}

@article{Mori2023,
  type      = {Journal {{Article}}},
  title     = {Floquet States in Open Quantum Systems},
  author    = {Mori, Takashi},
  year      = 2023,
  journal   = {Annu. Rev. Condens. Matter Phys.},
  volume    = {14},
  number    = {Volume 14, 2023},
  pages     = {35--56},
  publisher = {Annual Reviews},
  issn      = {1947-5462},
  doi       = {10.1146/annurev-conmatphys-040721-015537}
}

@article{Nakagawa2025,
  title     = {Topology of Discrete Quantum Feedback Control},
  author    = {Nakagawa, Masaya and Ueda, Masahito},
  journal   = {Phys. Rev. X},
  volume    = {15},
  issue     = {2},
  pages     = {021016},
  numpages  = {46},
  year      = {2025},
  month     = {Apr},
  publisher = {American Physical Society},
  doi       = {10.1103/PhysRevX.15.021016},
  url       = {https://link.aps.org/doi/10.1103/PhysRevX.15.021016}
}

@article{Oka2019,
  type      = {Journal {{Article}}},
  title     = {Floquet Engineering of Quantum Materials},
  author    = {Oka, Takashi and Kitamura, Sota},
  year      = 2019,
  journal   = {Annu. Rev. Condens. Matter Phys.},
  volume    = {10},
  number    = {Volume 10, 2019},
  pages     = {387--408},
  publisher = {Annual Reviews},
  issn      = {1947-5462},
  doi       = {10.1146/annurev-conmatphys-031218-013423}
}

@article{Okuma2020,
  title     = {Topological Origin of Non-Hermitian Skin Effects},
  author    = {Okuma, Nobuyuki and Kawabata, Kohei and Shiozaki, Ken and Sato, Masatoshi},
  year      = 2020,
  month     = feb,
  journal   = {Phys. Rev. Lett.},
  volume    = {124},
  number    = {8},
  pages     = {086801},
  publisher = {American Physical Society},
  doi       = {10.1103/PhysRevLett.124.086801}
}

@article{Okuma2023,
  type      = {Journal {{Article}}},
  title     = {Non-Hermitian Topological Phenomena: A Review},
  author    = {Okuma, Nobuyuki and Sato, Masatoshi},
  year      = 2023,
  journal   = {Annu. Rev. Condens. Matter Phys.},
  volume    = {14},
  number    = {Volume 14, 2023},
  pages     = {83--107},
  publisher = {Annual Reviews},
  issn      = {1947-5462},
  doi       = {10.1146/annurev-conmatphys-040521-033133}
}

@article{Pauli1927,
  title     = {Zur Quantenmechanik Des Magnetischen Elektrons},
  author    = {Pauli, W.},
  year      = 1927,
  month     = sep,
  journal   = {Z. F\"ur Phys.},
  volume    = {43},
  number    = {9--10},
  pages     = {601--623},
  publisher = {{Springer Science and Business Media LLC}},
  issn      = {0044-3328},
  doi       = {10.1007/bf01397326}
}

@article{Pavlov2025,
  title     = {Topological transitions in quantum jump dynamics: Hidden exceptional points},
  author    = {Pavlov, Andrei I. and Gefen, Yuval and Shnirman, Alexander},
  journal   = {Phys. Rev. B},
  volume    = {111},
  issue     = {10},
  pages     = {104301},
  numpages  = {19},
  year      = {2025},
  month     = {Mar},
  publisher = {American Physical Society},
  doi       = {10.1103/PhysRevB.111.104301},
  url       = {https://link.aps.org/doi/10.1103/PhysRevB.111.104301}
}

@article{Potter2016,
  title     = {Classification of Interacting Topological Floquet Phases in One Dimension},
  author    = {Potter, Andrew C. and Morimoto, Takahiro and Vishwanath, Ashvin},
  year      = 2016,
  month     = oct,
  journal   = {Phys. Rev. X},
  volume    = {6},
  number    = {4},
  pages     = {041001},
  publisher = {American Physical Society},
  doi       = {10.1103/PhysRevX.6.041001}
}

@article{Quelle2017,
  title     = {Driving Protocol for a Floquet Topological Phase without Static Counterpart},
  author    = {Quelle, A and Weitenberg, C and Sengstock, K and Smith, C Morais},
  year      = 2017,
  month     = nov,
  journal   = {New J. Phys.},
  volume    = {19},
  number    = {11},
  pages     = {113010},
  publisher = {IOP Publishing},
  doi       = {10.1088/1367-2630/aa8646}
}

@article{Rudner2013,
  title     = {Anomalous Edge States and the Bulk-Edge Correspondence for Periodically Driven Two-Dimensional Systems},
  author    = {Rudner, Mark S. and Lindner, Netanel H. and Berg, Erez and Levin, Michael},
  year      = 2013,
  month     = jul,
  journal   = {Phys. Rev. X},
  volume    = {3},
  number    = {3},
  pages     = {031005},
  publisher = {American Physical Society},
  doi       = {10.1103/PhysRevX.3.031005}
}

@article{Schnell2020,
  title     = {Is there a Floquet Lindbladian?},
  author    = {Schnell, Alexander and Eckardt, Andr\'e and Denisov, Sergey},
  journal   = {Phys. Rev. B},
  volume    = {101},
  issue     = {10},
  pages     = {100301},
  numpages  = {6},
  year      = {2020},
  month     = {Mar},
  publisher = {American Physical Society},
  doi       = {10.1103/PhysRevB.101.100301},
  url       = {https://link.aps.org/doi/10.1103/PhysRevB.101.100301}
}

@article{Sergi2019,
  title    = {The Density Matrix in the Non-{{Hermitian}} Approach to Open Quantum System Dynamics},
  author   = {Sergi, Alessandro},
  year     = 2019,
  month    = dec,
  journal  = {Atti Accad. Peloritana Dei Pericolanti Cl. Sci. Fis. Mat. Nat.},
  volume   = {97},
  number   = {S2},
  pages    = {11}
}

@article{Sierant2022,
  title     = {Dissipative Floquet Dynamics: From Steady State to Measurement Induced Criticality in Trapped-Ion Chains},
  author    = {Sierant, Piotr and Chiriac{\`o}, Giuliano and Surace, Federica M. and Sharma, Shraddha and Turkeshi, Xhek and Dalmonte, Marcello and Fazio, Rosario and Pagano, Guido},
  year      = 2022,
  month     = feb,
  journal   = {Quantum},
  volume    = {6},
  pages     = {638},
  publisher = {Verein zur F\"orderung des Open Access Publizierens in den Quantenwissenschaften},
  issn      = {2521-327X},
  doi       = {10.22331/q-2022-02-02-638}
}

@article{Song2019,
  title     = {Non-Hermitian Skin Effect and Chiral Damping in Open Quantum Systems},
  author    = {Song, Fei and Yao, Shunyu and Wang, Zhong},
  year      = 2019,
  month     = oct,
  journal   = {Phys. Rev. Lett.},
  volume    = {123},
  number    = {17},
  pages     = {170401},
  publisher = {American Physical Society},
  doi       = {10.1103/PhysRevLett.123.170401}
}

@article{Tao2013,
  title   = {Outliers in the Spectrum of Iid Matrices with Bounded Rank Perturbations},
  author  = {Tao, Terence},
  year    = 2013,
  month   = feb,
  journal = {Probab. Theory Relat. Fields},
  volume  = {155},
  number  = {1-2},
  pages   = {231--263},
  issn    = {0178-8051, 1432-2064},
  doi     = {10.1007/s00440-011-0397-9},
  urldate = {2025-11-13},
  langid  = {english}
}

@article{Vajna2015,
  title     = {Topological Classification of Dynamical Phase Transitions},
  author    = {Vajna, Szabolcs and D{\'o}ra, Bal{\'a}zs},
  year      = 2015,
  month     = apr,
  journal   = {Phys. Rev. B},
  volume    = {91},
  number    = {15},
  pages     = {155127},
  publisher = {American Physical Society},
  doi       = {10.1103/PhysRevB.91.155127}
}

@misc{Wanjura2025,
  title         = {Unifying framework for non-Hermitian and Hermitian topology in driven-dissipative systems},
  author        = {Clara C. Wanjura and Andreas Nunnenkamp},
  year          = {2025},
  eprint        = {2509.19433},
  archiveprefix = {arXiv},
  primaryclass  = {cond-mat.mes-hall},
  url           = {https://arxiv.org/abs/2509.19433}
}

@article{Wiseman1996,
  title     = {Quantum Trajectories and Quantum Measurement Theory},
  author    = {Wiseman, H M},
  year      = 1996,
  month     = feb,
  journal   = {Quantum Semiclassical Opt. J. Eur. Opt. Soc. Part B},
  volume    = {8},
  number    = {1},
  pages     = {205--222},
  publisher = {IOP Publishing},
  issn      = {1361-6625},
  doi       = {10.1088/1355-5111/8/1/015}
}

@article{Wu2021,
  title     = {Floquet Second-Order Topological Insulators in Non-{{Hermitian}} Systems},
  author    = {Wu, Hong and Wang, Bao-Qin and An, Jun-Hong},
  year      = 2021,
  month     = jan,
  journal   = {Phys. Rev. B},
  volume    = {103},
  number    = {4},
  pages     = {L041115},
  publisher = {American Physical Society},
  doi       = {10.1103/PhysRevB.103.L041115}
}

@article{Yao2017,
  title     = {Topological Invariants of Floquet Systems: {{General}} Formulation, Special Properties, and Floquet Topological Defects},
  author    = {Yao, Shunyu and Yan, Zhongbo and Wang, Zhong},
  year      = 2017,
  month     = nov,
  journal   = {Phys. Rev. B},
  volume    = {96},
  number    = {19},
  pages     = {195303},
  publisher = {American Physical Society},
  doi       = {10.1103/PhysRevB.96.195303}
}

@article{Yao2018,
  title     = {Edge States and Topological Invariants of Non-Hermitian Systems},
  author    = {Yao, Shunyu and Wang, Zhong},
  year      = 2018,
  month     = aug,
  journal   = {Phys. Rev. Lett.},
  volume    = {121},
  number    = {8},
  pages     = {086803},
  publisher = {American Physical Society},
  doi       = {10.1103/PhysRevLett.121.086803}
}

@article{Yokomizo2019,
  title     = {Non-Bloch Band Theory of Non-Hermitian Systems},
  author    = {Yokomizo, Kazuki and Murakami, Shuichi},
  year      = 2019,
  month     = aug,
  journal   = {Phys. Rev. Lett.},
  volume    = {123},
  number    = {6},
  pages     = {066404},
  publisher = {American Physical Society},
  doi       = {10.1103/PhysRevLett.123.066404}
}

@misc{Zenodo2026,
  title     = {Data set for ``Liouvillian topology and nonreciprocal dynamics in open Floquet chains''},
  author    = {Koch, Florian and Budich, Jan Carl and Hu, Yu-Min},
  publisher = {Zenodo},
  year      = {2026},
  doi       = {10.5281/zenodo.20325478}
}

@article{Zhang2020,
  title     = {Correspondence between Winding Numbers and Skin Modes in Non-Hermitian Systems},
  author    = {Zhang, Kai and Yang, Zhesen and Fang, Chen},
  year      = 2020,
  month     = sep,
  journal   = {Phys. Rev. Lett.},
  volume    = {125},
  number    = {12},
  pages     = {126402},
  publisher = {American Physical Society},
  doi       = {10.1103/PhysRevLett.125.126402}
}

@article{Zhang2020a,
  title     = {Non-Hermitian Floquet Topological Phases: {{Exceptional}} Points, Coalescent Edge Modes, and the Skin Effect},
  author    = {Zhang, Xizheng and Gong, Jiangbin},
  year      = 2020,
  journal   = {Phys. Rev. B},
  volume    = {101},
  number    = {4},
  pages     = {045415},
  publisher = {APS}
}

@article{Zhou2019,
  title     = {Periodic Table for Topological Bands with Non-{{Hermitian}} Symmetries},
  author    = {Zhou, Hengyun and Lee, Jong Yeon},
  year      = 2019,
  month     = jun,
  journal   = {Phys. Rev. B},
  volume    = {99},
  number    = {23},
  pages     = {235112},
  publisher = {American Physical Society},
  doi       = {10.1103/PhysRevB.99.235112}
}

@article{Zhou2023,
  title    = {Non-Hermitian Floquet Topological Matter---a Review},
  author   = {Zhou, Longwen and Zhang, Da-Jian},
  year     = 2023,
  journal  = {Entropy},
  volume   = {25},
  number   = {10},
  issn     = {1099-4300},
  doi      = {10.3390/e25101401}
}


\end{document}